\documentclass[aps,pra,twocolumn,superscriptaddress,amsmath,amssymb,floatfix]{revtex4}
\usepackage{color}
\usepackage{graphicx}
\usepackage{bm}
\usepackage{mathrsfs}

\begin{document}

\author{Marko Gacesa}
\email{marko.gacesa@nasa.gov}

\affiliation{NASA Ames Research Center, Moffett Field, CA, USA}
\affiliation{Department of Physics, University of Connecticut, Storrs, CT, USA}

\author{John A. Montgomery, Jr.}
\affiliation{Department of Physics, University of Connecticut, Storrs, CT, USA}

\author{H. Harvey Michels}
\affiliation{Department of Physics, University of Connecticut, Storrs, CT, USA}

\author{Robin C\^ot\'e}
\affiliation{Department of Physics, University of Connecticut, Storrs, CT, USA}

\title{Production of NaCa$^+$ molecular ions in the ground state from cold atom-ion mixtures by photoassociation via an intermediate state}

\date{\today}

\begin{abstract} 
We present a theoretical analysis of optical pathways for formation of cold ground state (NaCa)$^+$ molecular ions via an intermediate state. The formation schemes are based on \textit{ab initio} potential energy curves and transition dipole moments calculated using effective-core-potential methods of quantum chemistry. 
In the proposed approach, starting from a mixture of cold trapped Ca$^+$ ions immersed into an ultracold gas of Na atoms, (NaCa)$^+$ molecular ions are photoassociated in the excited E$^{1}\Sigma^+$ electronic state and allowed to spontaneously decay either to the ground electronic state or an intermediate state from which the population is transferred to the ground state via an additional optical excitation.
By analyzing all possible pathways, we find that the efficiency of a two-photon scheme, via either B$^{1}\Sigma^+$ or C$^{1}\Sigma^+$ potential, is sufficient to produce significant quantities of ground state (NaCa)$^+$ molecular ions. A single-step process results in lower formation rates that would require either a high density sample or a very intense photoassociation laser to be viable.
\end{abstract}

\maketitle

\section{Introduction}
\label{sec:introduction}

At present, there exists a strong interest in the scientific community in techniques that reliably produce large samples of ultracold molecules of different species. Since direct laser cooling of molecules is impractical due to their internal rotational and vibrational (ro-vibrational) degrees of freedom, a number of different approaches that allow production of ultracold samples of selected dimers have been developed\cite{krems2009cold,2009NJPh...11e5049C}. 
Recent advances in trapping and laser-cooling ions to ultracold temperatures\cite{2005PhRvL..95r3002B,2009PhRvA..79c2716H,2010NaPho...4..772S,2010NatPh...6..275S,2014NatCo...5E5587H,2015PhRvA..91b3430K,2016PhRvA..93a1401L} allow experimenting with hybrid systems composed of overlapping trapped cold atomic gases with ultracold ions\cite{2014ConPh..55...33H}. 
Such systems offer opportunities for new developments in the field of ultracold quantum matter, with a benefit of simpler and more reliable trapping than it is available for neutral molecules.

For example, hybrid atom-ion systems have been proposed as emulators of periodic condensed matter systems with a band structure\cite{2013PhRvL.111h0501B}, a possible implementation of an atom-atom Josephson junction\cite{2012PhRvL.109h0402G}, and as platforms for studying charge mobility in an ultracold gas\cite{2000PhRvL..85.5316C}. Ions immersed in a Bose-Einstein condensate were proposed as a medium for exploring the physics of mesoscopic particles\cite{2002PhRvL..89i3001C}, polarons\cite{2006PhRvL..96u0401C}, and electron-phonon coupling\cite{2013Natur.502..664B,2015PhRvL.114x3003W}. 
Hybrid atom-ions systems are also discussed as possible implementations of quantum gates\cite{2010PhRvA..81a2708D,2010PhRvA..81c0301K}.
In addition, trapped samples of cold molecular ions allow investigations of collisional dynamics and chemistry at temperatures of the order of millikelvin or below, where the long-range nature of atom-ion interaction is responsible for qualitative differences from neutral molecules\cite{2000PhRvA..62a2709C,2011PhRvA..83f2712G,2013PhRvA..88b2701G,2015JChPh.143d1105P}. 
These studies are also a key to identifying efficient and versatile ways to produce samples of ground state molecular ions, required for majority of proposed applications, for species where sympathetic cooling with ultracold atoms is not applicable or sufficiently effective\cite{2012PhRvL.109y3201C,2012NatCo...3E1126R,2015PhRvA..91b3430K}.

Reactive processes of particular interest are radiative association (RA) from the continuum and radiative charge-exchange (RCX). These processes are often in competition with each other and commonly analyzed together. The RA has been observed for Rb+Ca$^+$\cite{2011PhRvL.107x3202H,2013MolPh.111.2020H} and Rb+Ba$^+$\cite{2013MolPh.111.1683H}, and both processes have been investigated for several molecular ion species, including H+(D$^+$,T$^+$) and D+T$^+$\cite{2008NJPh...10c3024B}, Rb+Na$^+$\cite{2013PhRvA..88a2709Y,2014PhRvA..90c2714Y}, Li+Yb$^+$\cite{2015PhRvA..91d2706T}, Rb+Yb$^+$\cite{2013PhRvA..87e2717S}, as well as for Rb+(Ca$^+$,Sr$^+$,Ba$^+$,Yb$^+$), Li+Yb$^+$\cite{2015NJPh...17d5015D}, Li+Be$^+$\cite{2011PhRvA..83b2703R}, and possibly in Ca+Yb$^{+}$\cite{2011PhRvL.107x3201R,2014JPhB...47a5301Z}. The experiments and theoretical studies with emphasis on radiative charge-exchange in atom-ion systems have been conducted for Na+Na$^+$\cite{2000PhRvA..62a2709C,2015PhRvA..91a2709G}, Na+Ca$^+$\cite{2003PhRvA..67d2705M,2014ApPhB.114...75S}, Be+Be$^+$\cite{2011PCCP...1319026Z}, Yb+Yb$^+$\cite{2009PhRvL.102v3201G}, Ca+Yb$^+$ and Ca+Ba$^+$, Rb+Yb$^+$\cite{2014JPhB...47n5201M}, Na+Li$^+$\cite{2015PhRvA..91e2702L}, and others\cite{2015NJPh...17d5015D,2015arXiv150403114R}. See Ref. \cite{2014ConPh..55...33H} for a more extensive review.

\begin{figure}[t]
 \centering
 \includegraphics[clip,width=\linewidth]{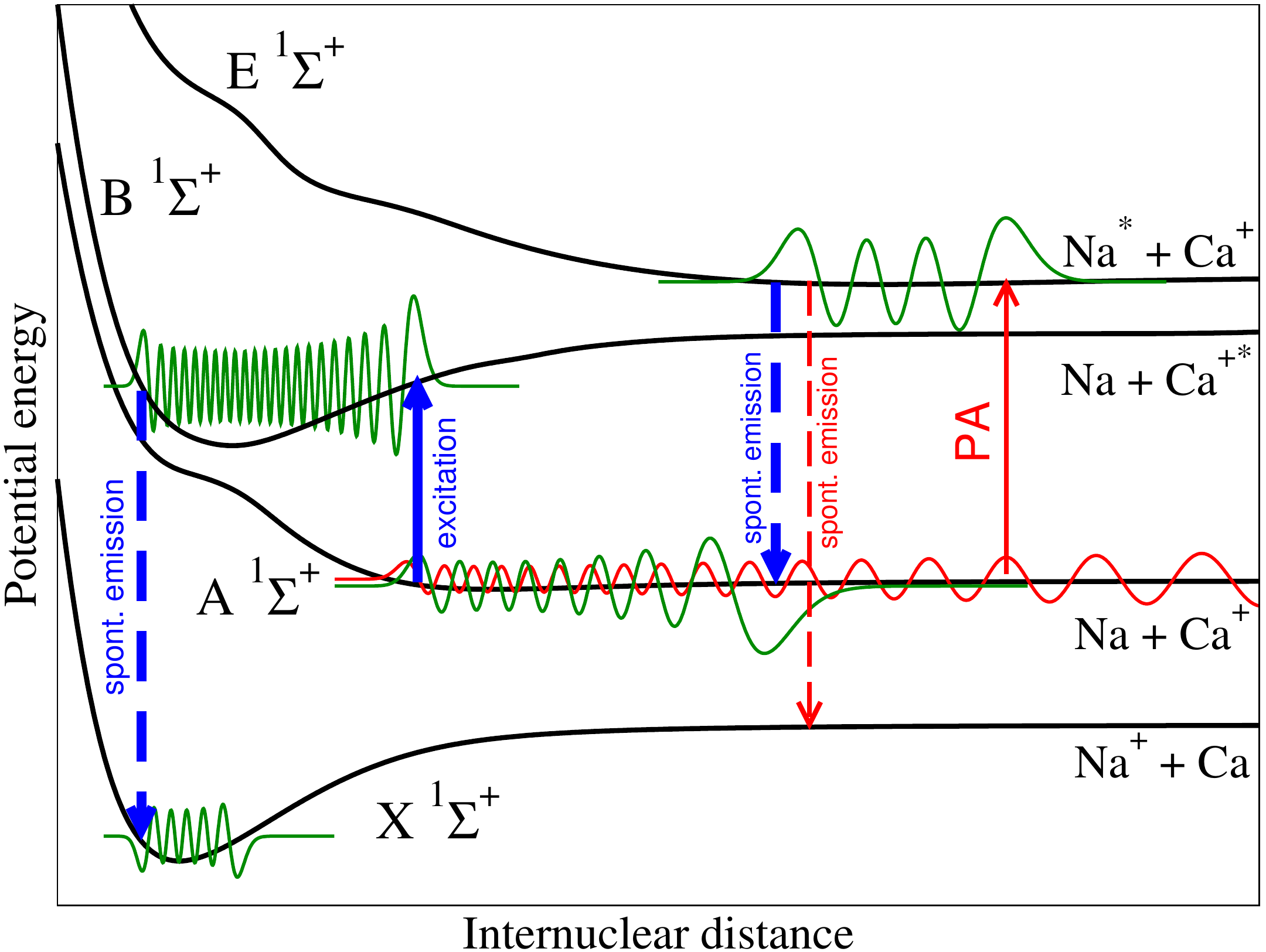}
 \caption{(Color online) Schematic representation of the studied optical pathways to produce NaCa$^+$ molecular ions in the electronic ground state.}
 \label{fig:scheme}
\end{figure}

In this article, we present a theoretically study of production of ultracold sample of NaCa$^+$ molecular ions by photoassociation (PA) in a mixture of ultracold Na atoms immersed in trapped laser-cooled Ca$^+$ ions. We investigate an optical scheme (Fig. \ref{fig:scheme}) where NaCa$^+$ molecular ions are photoassociated in an excited electronic state and allowed to spontaneously decay into the ground state in one or more steps, followed by a series of ro-vibrational transitions until settling in the lowest vibrational level. By analyzing all relevant optical transitions and allowing for a second optical excitation to an intermediate excited state to increase the efficiency of the processs, we found an optimal set of pathways that could lead to efficient production of ultracold NaCa$^+$ ions.

Our approach is motivated by the fact that the PA is a highly successful and state-selective method of production of cold diatomic molecules from ultracold atomic gases\cite{1987PhRvL..58.2420T,RevModPhys.78.483,stwalley1999photoassociation,krems2009cold} and could be considered a possible alternative to RA or RCX in molecular ions with favorable electronic structure\cite{2011PhRvA..83b2703R}. 
Our choice of the system was motivated by ongoing experiments\cite{2014ApPhB.114...75S}, existing theoretical results\cite{2003PhRvA..67d2705M}, and the fact that similar electronic configuration of the outer shell is shared by other molecular ion species currently of interest\cite{2015NJPh...17d5015D}.

The article is organized as follows. In Section \ref{sec:methods} we give a brief overview of the theoretical methods used in this study to calculate the photoassociation rates and spontaneous emission rates to lower electronic states. The details of \textit{ab initio} calculations of the electronic structure are given and calculated electronic potentials and dipole moments are presented. In Section \ref{sec:results} we present main results of this study, including dipole transition matrix elements for relevant pairs of electronic states involved in the optical production pathways, photoassociation and relaxation rates, and a description of vibrational relaxation in the ground state. The conclusions are given in Section \ref{sec:summary}.

\section{Theoretical methods}
\label{sec:methods}


\subsection{Electronic structure calculation}
\label{subsec:th_structure}

The potential energy curves and transition dipole moments for the ground and low-lying excited electronic states of singlet symmetry were calculated using the equation-of-motion coupled cluster singles and doubles method (EOM-CCSD)\cite{Stanton_Bartlett_1993,2003JChPh.118.3006K}, as implemented in MOLPRO program package\cite{MOLPRO}. 
Inner shell electrons were replaced by an effective core potential (ECP), while core-valence effects were included using a core polarization potential\cite{Fuentealba1982,Fuentealba1985}.
Extended basis sets described in detail in Refs. \cite{2005JChPh.122t4302A} and \cite{czuchaj2003valence} were used to describe the two valence electrons. Specifically, the exponents of Gaussian functions (set B) with uncontracted orbitals) from Ref. \cite{2005JChPh.122t4302A} were used for Na atomic basis, while a contracted and optimized\cite{Fuentealba1985} basis set for Ca, augmented by three sets of $f$ polarization functions and a single $g$ function\cite{czuchaj2003valence} was used for atomic Ca. 
With these approximations, our approach is equivalent to performing a full valence configuration interaction (CI) calculation. A similar approach proved highly accurate in our study of Ca$_{2}^{+}$\cite{2012CPL...542..138B}. We note that the inclusion of ECP and CPP is critical for obtaining correct asymptote ordering.
The potentials were interpolated using a cubic spline. 

\begin{figure}[t]
 \centering
 \includegraphics[clip,width=\linewidth]{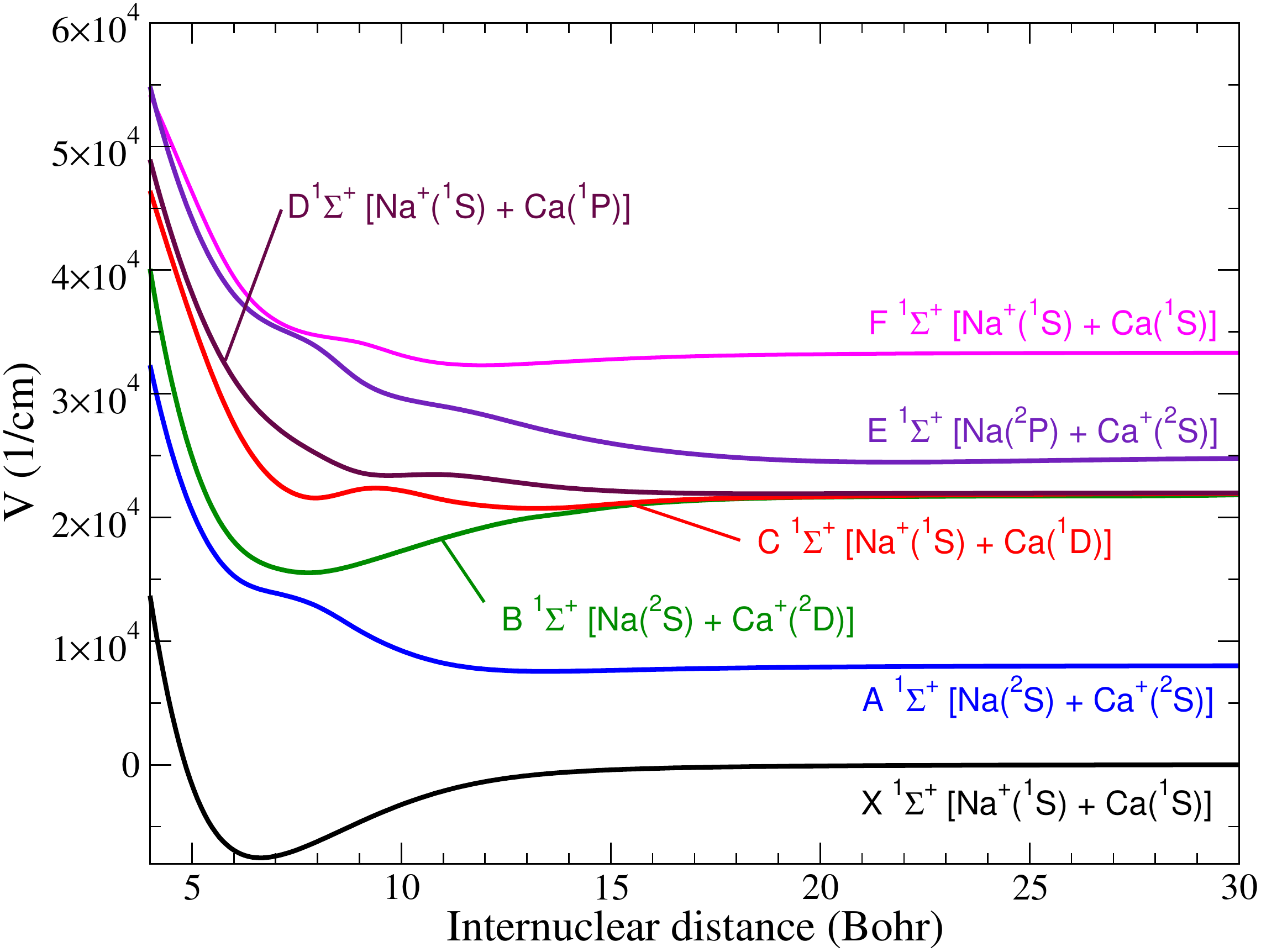}
 \caption{(Color online) Calculated NaCa$^+$ potential energy curves of $^{1}\Sigma^{+}$ symmetry.}
 \label{fig:potentials_sigma}
\end{figure}
\begin{figure}[t]
 \centering
 \includegraphics[clip,width=\linewidth]{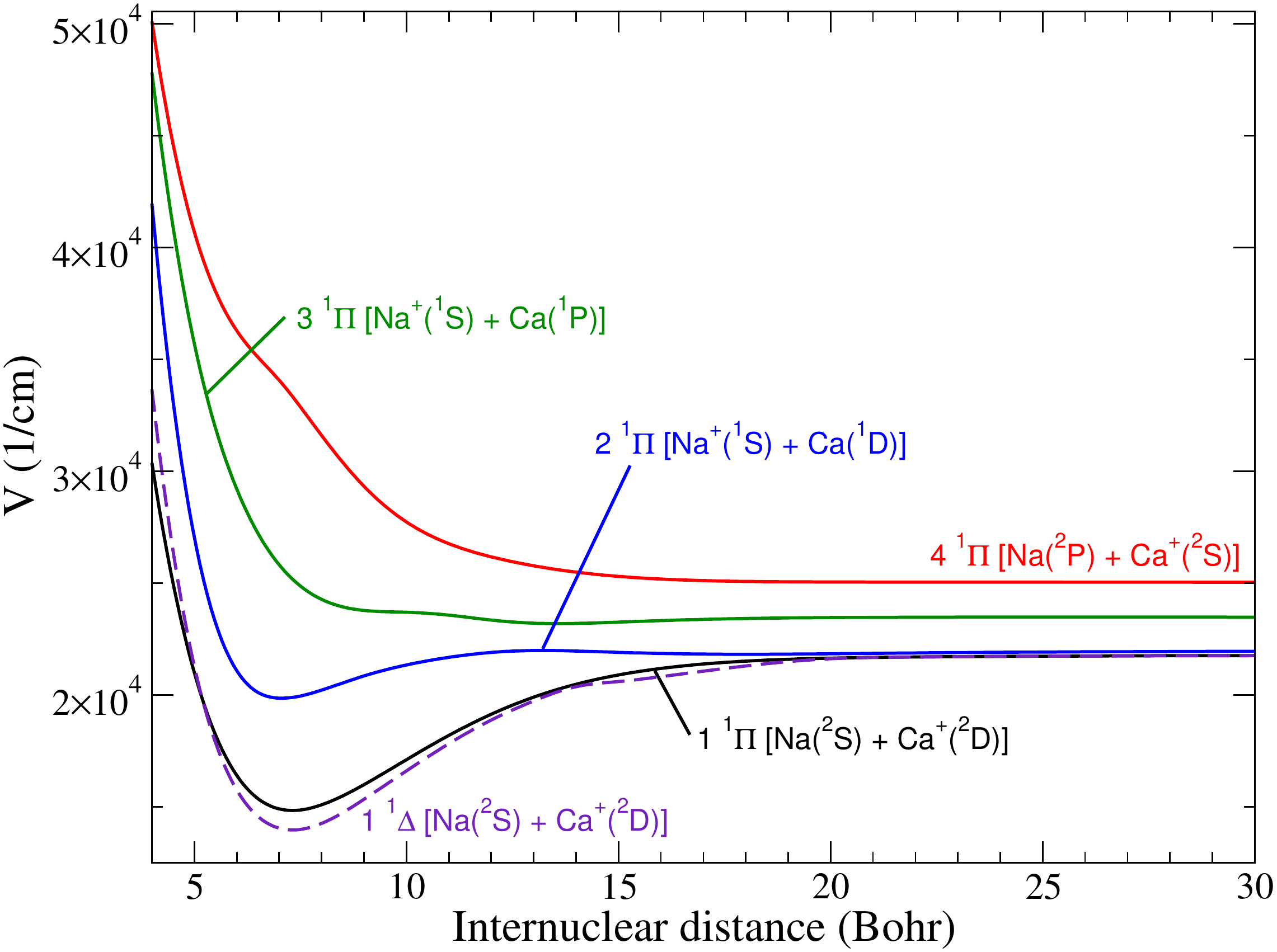}
 \caption{(Color online) Calculated NaCa$^+$ potential energy curves of $^{1}\Pi$ and $^{1}\Delta$ symmetry. }
 \label{fig:potentials_pi}
\end{figure}

Calculated adiabatic potential energy curves are given in Figs. \ref{fig:potentials_sigma} and \ref{fig:potentials_pi}. 
Our calculation gives a deeper X$^{1}\Sigma^{+}$ potential and a significantly shallower well in A$^{1}\Sigma^{+}$ with a qualitatively different inner wall than an earlier result obtained using second order M\"oller-Plesset perturbation theory (MP2) with a Gaussian triple $\zeta$+diffuse+polarization basis set, 6-311+G(3df)\cite{2003PhRvA..67d2705M}.
The calculated \textit{ab-initio} points are given in Tables \ref{t:pots_sigma} and \ref{t:pots_pi}.

\begin{table*}
\caption{Calculated ab-initio points of the potential energy curves of $^{1}\Sigma^+$ symmetry (in Hartree).}
\begin{ruledtabular}
\begin{tabular}{rcccccc}
R(Bohr) & X$^1\Sigma^+$ & A$^1\Sigma^+$ & B$^1\Sigma^+$ & C$^1\Sigma^+$ & D$^1\Sigma^+$  & E$^1\Sigma^+$  \\
\hline
4&	-0.599284&	-0.514354&	-0.478884&	-0.450140&	-0.438636&	-0.411727 \\
5&	-0.668878&	-0.567910&	-0.548060&	-0.497608&	-0.487797&	-0.460320 \\
6&	-0.693037&	-0.591946&	-0.579284&	-0.535934&	-0.519360&	-0.488312 \\
7&	-0.695231&	-0.598355&	-0.589118&	-0.557372&	-0.536956&	-0.500249 \\
8&	-0.689814&	-0.603266&	-0.590707&	-0.563287&	-0.547163&	-0.507788 \\
9&	-0.682733&	-0.612139&	-0.587537&	-0.560104&	-0.553874&	-0.519958 \\
10&	-0.676279&	-0.619522&	-0.582831&	-0.560566&	-0.554916&	-0.526541 \\
11&	-0.671262&	-0.624112&	-0.578106&	-0.563934&	-0.554656&	-0.529511 \\
12&	-0.667794&	-0.626364&	-0.573985&	-0.566150&	-0.556043&	-0.532744 \\
13&	-0.665597&	-0.627111&	-0.570776&	-0.567103&	-0.558019&	-0.536675 \\
14&	-0.664264&	-0.627101&	-0.568754&	-0.566773&	-0.559638&	-0.540250 \\
15&	-0.663448&	-0.626809&	-0.566644&	-0.565568&	-0.560645&	-0.543160 \\
16&	-0.662926&	-0.626468&	-0.565188&	-0.564494&	-0.561238&	-0.545423 \\
17&	-0.662577&	-0.626165&	-0.564220&	-0.563767&	-0.561557&	-0.547129 \\
18&	-0.662332&	-0.625921&	-0.563600&	-0.563316&	-0.561695&	-0.548370 \\
19&	-0.662154&	-0.625733&	-0.563218&	-0.563037&	-0.561722&	-0.549223 \\
20&	-0.662022&	-0.625589&	-0.562985&	-0.562830&	-0.561693&	-0.549753 \\
22&	-0.661844&	-0.625394&	-0.562725&	-0.562312&	-0.561597&	-0.550073 \\
24&	-0.661734&	-0.625275&	-0.562603&	-0.562004&	-0.561522&	-0.549805 \\
26&	-0.661662&	-0.625200&	-0.562529&	-0.561834&	-0.561476&	-0.549373 \\
28&	-0.661615&	-0.625149&	-0.562481&	-0.561728&	-0.561449&	-0.548987 \\
30&	-0.661582&	-0.625115&	-0.562103&	-0.561657&	-0.561432&	-0.548693
\end{tabular}
\end{ruledtabular}
\label{t:pots_sigma}
\end{table*}

\begin{table*}
\caption{As above for $^{1}\Pi$ and $^{1}\Delta$ symmetry.}
\begin{ruledtabular}
\begin{tabular}{rccccc}
R(Bohr)  & $1^1\Pi$   & $2^1\Pi$        & $3^1\Pi$     & $4^1\Pi$ &      $1^1\Delta$  \\
\hline
4&	-0.523201&	-0.470398&	-0.443687&	-0.433276 &	 -0.508242\\
5&	-0.565926&	-0.538373&	-0.498950&	-0.476216 &	 -0.564629\\
6&	-0.586845&	-0.565520&	-0.528540&	-0.496293 &	 -0.589801\\
7&	-0.593672&	-0.571116&	-0.543963&	-0.506382 &	 -0.597589\\
8&	-0.592831&	-0.569479&	-0.550995&	-0.517620 &	 -0.596621\\
9&	-0.588726&	-0.566593&	-0.553259&	-0.527986 &	 -0.591701\\
10&	-0.583653&	-0.564361&	-0.553623&	-0.535240 &	 -0.585930\\
11&	-0.578702&	-0.562806&	-0.554218&	-0.539662 &	 -0.580385\\
12&	-0.574388&	-0.561806&	-0.555215&	-0.542344 &	 -0.575570\\
13&	-0.570913&	-0.561419&	-0.555851&	-0.544166 &	 -0.571689\\
14&	-0.568285&	-0.561522&	-0.555895&	-0.545452 &	 -0.568779\\
15&	-0.566389&	-0.561789&	-0.555635&	-0.546321 &	 -0.567744\\
16&	-0.565065&	-0.562012&	-0.555328&	-0.546867 &	 -0.566739\\
17&	-0.564167&	-0.562140&	-0.555077&	-0.547183 &	 -0.565626\\
18&	-0.563576&	-0.562178&	-0.554899&	-0.547349 &	 -0.564557\\
19&	-0.563201&	-0.562148&	-0.554782&	-0.547429 &	 -0.563687\\
20&	-0.562969&	-0.562076&	-0.554710&	-0.547464 &	 -0.563096\\
22&	-0.562727&	-0.561909&	-0.554640&	-0.547484 &	 -0.562738\\
24&	-0.562605&	-0.561778&	-0.554619&	-0.547491 &	 -0.562612\\
26&	-0.562531&	-0.561687&	-0.554617&	-0.547502 &	 -0.562535\\
28&	-0.562481&	-0.561624&	-0.554622&	-0.547515 &	 -0.562484\\
30&	-0.562448&	-0.561579&	-0.554629&	-0.547530 &	 -0.562447
\end{tabular}
\end{ruledtabular}
\label{t:pots_pi}
\end{table*}

\begin{figure}[t]
 \centering
 \includegraphics[clip,width=\linewidth]{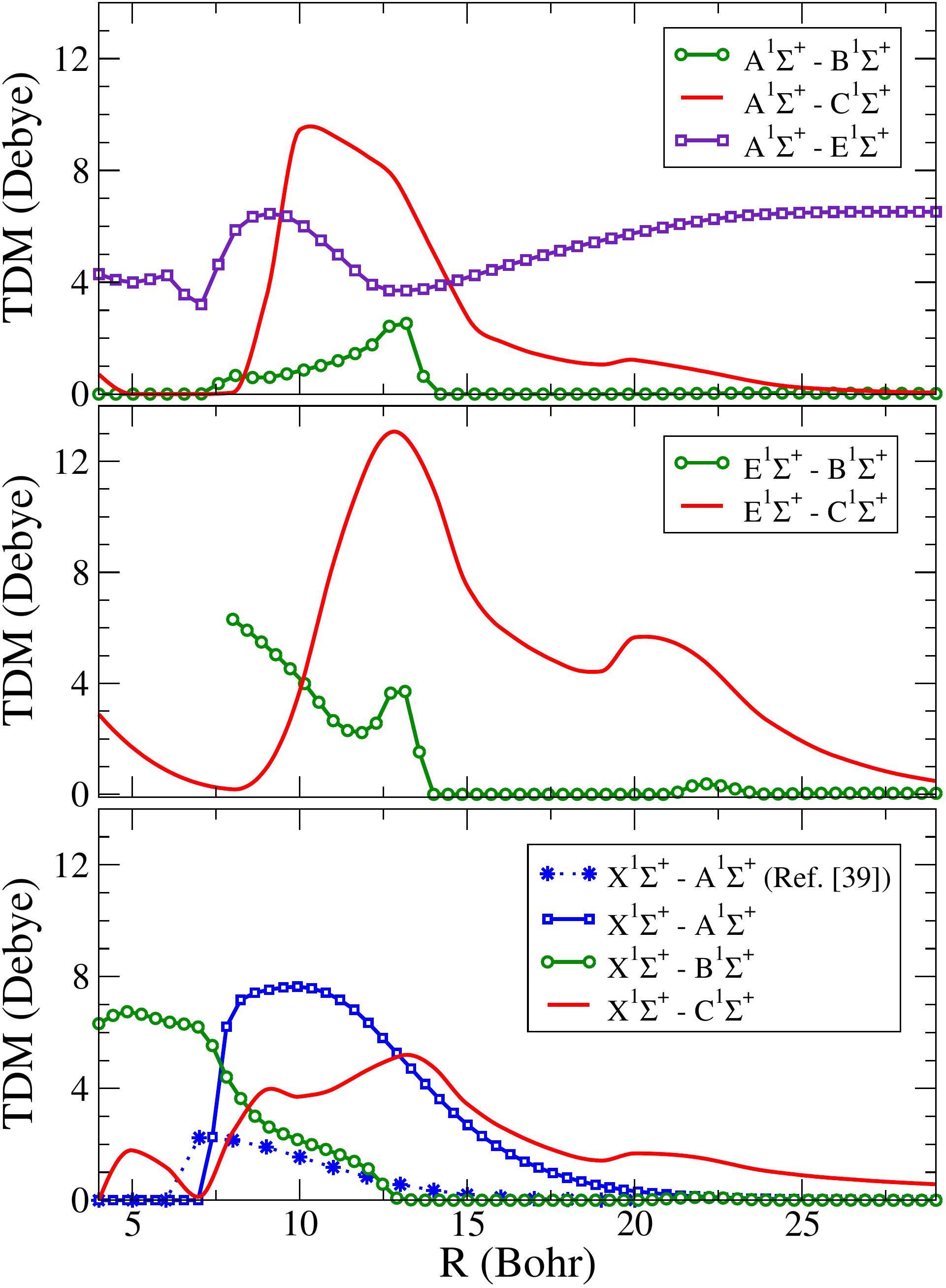}
 \caption{(Color online) Transition dipole moments (TDM) between selected electronic states for $A^{1}\Sigma^{+}$ (top), $E^{1}\Sigma^{+}$ (middle), and $X^{1}\Sigma^{+}$ (bottom) as functions of internuclear distance. The result from Ref. \cite{2003PhRvA..67d2705M} is also shown (bottom).}
 \label{fig:dipolemoments}
\end{figure}

\begin{figure}[t]
 \centering
 \includegraphics[clip,width=\linewidth]{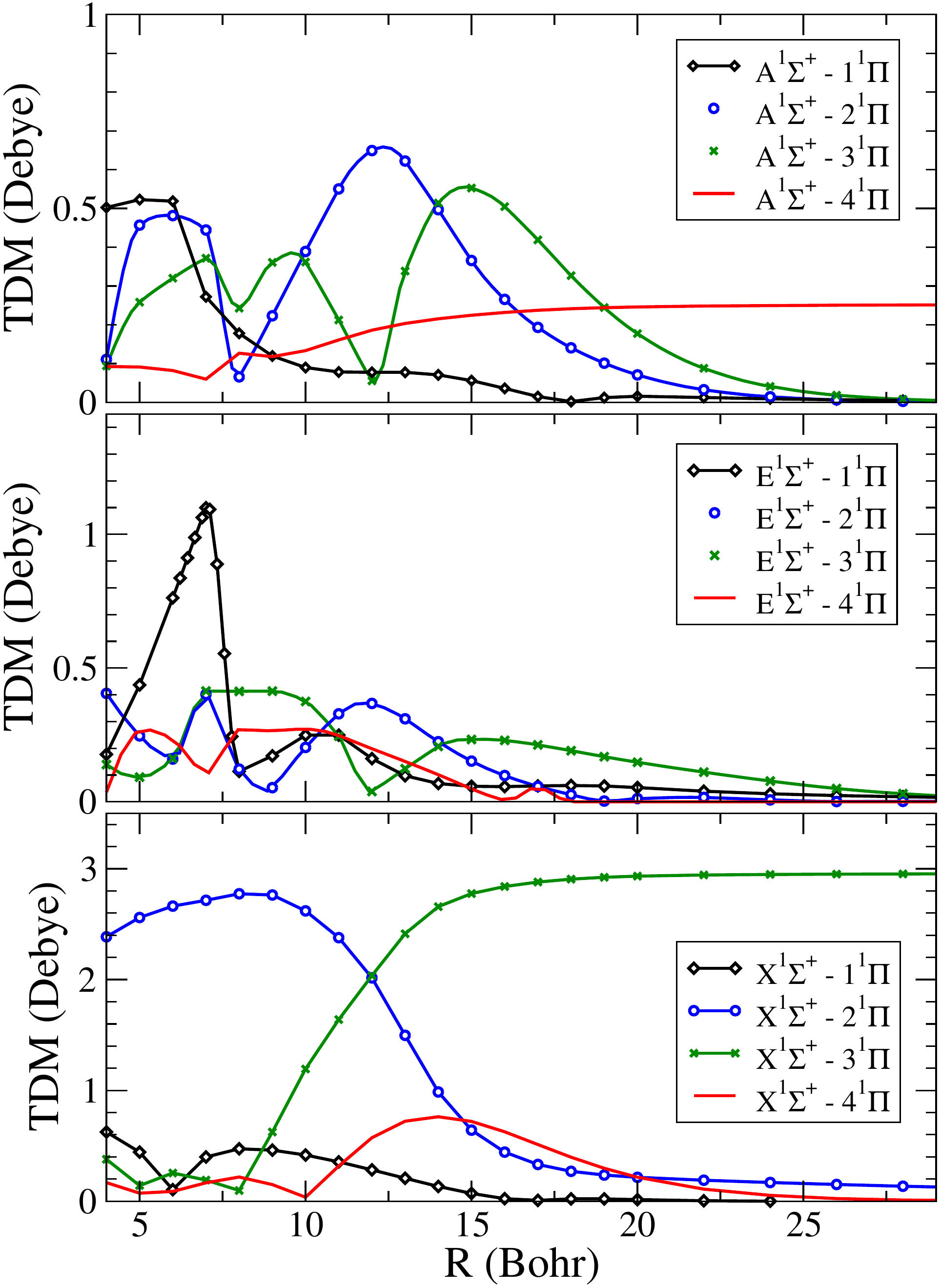}
 \caption{(Color online) As Fig. \ref{fig:dipolemoments} for selected $^1\Sigma^{+} - ^1\Pi$ transitions.}
 \label{fig:dipolemoments-pi}
\end{figure}

\begin{figure}[t]
 \centering
 \includegraphics[clip,width=\linewidth]{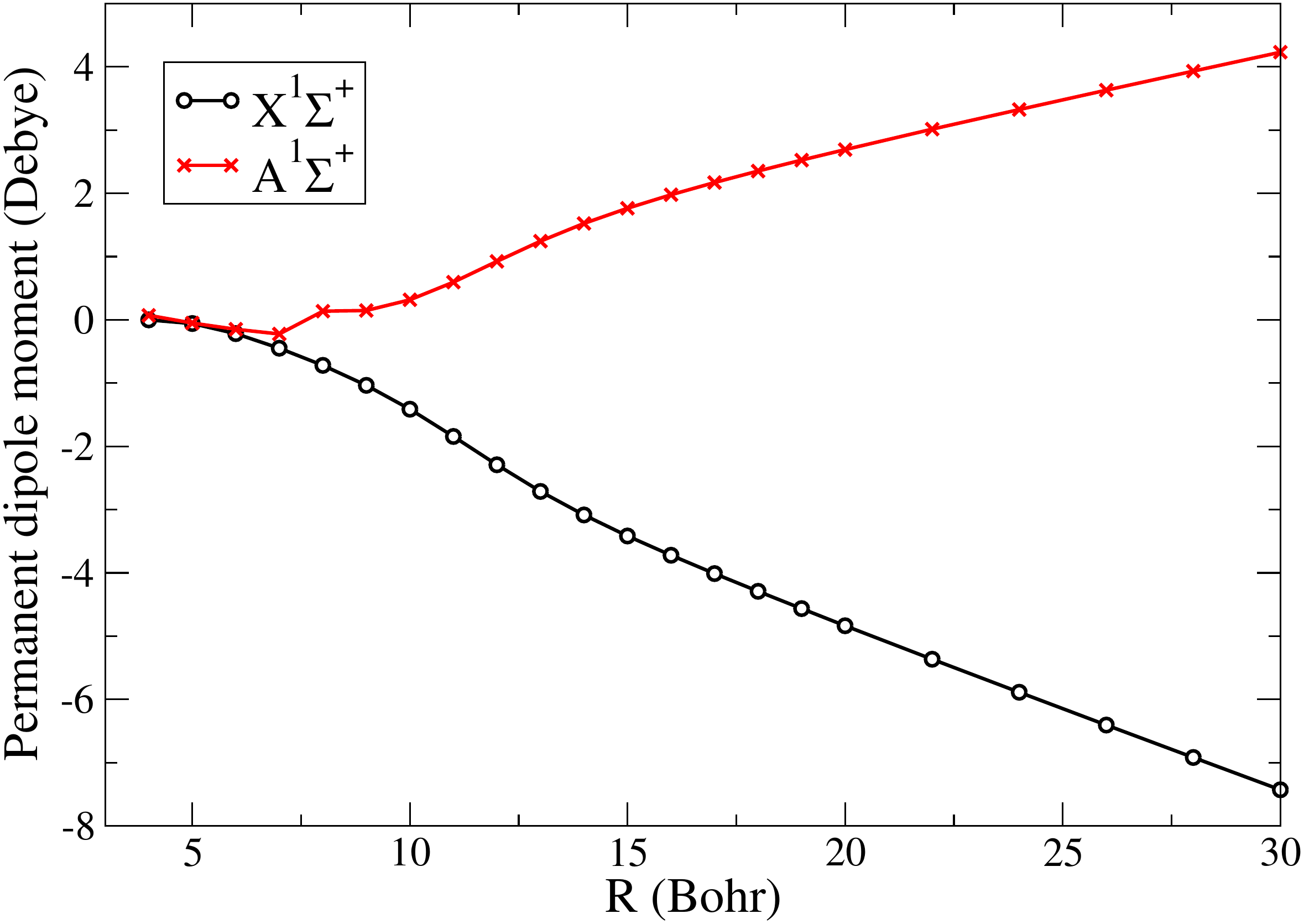}
 \caption{(Color online) Permanent dipole moments for $X^{1}\Sigma^{+}$ and $A^{1}\Sigma^{+}$ electronic states of NaCa$^+$ as functions of internuclear distance. Calculated \textit{ab initio} points are shown.}
 \label{fig:permdipolemoments}
\end{figure}

The long-range ($R>30$ Bohr) part of the interaction potentials was constructed using the polarizabilities and dispersion coefficients calculated by Kaur \textit{et al.}\cite{2015PhRvA..91a2705K}, and connected smoothly to the \textit{ab initio} points. 
Specifically, for the A$^1\Sigma^+$ state, where the dispersion coefficients are known, the long-range form used was\begin{equation}
  V_{\mathrm{lr}}(R) = - \frac{C_4}{R^4} - \frac{C_6}{R^6} - \frac{C_8}{R^8} - \frac{C_{10}}{R^{10}} ,
\end{equation}
with $C_4 = 81.2$, $C_6 = 901$, $C_8 = 60720$, and $C_{10} = 5.4729 \times 10^6$ (from Ref. \cite{2015PhRvA..91a2705K}, in atomic units). For the ground state and other excited states for which the dispersion coefficients are not well known, the long-range was described using
\begin{equation}
  V_{\mathrm{lr}}(R) = - \frac{1}{2} \frac{\alpha_d}{R^4} ,
\end{equation}
where the static dipole polarizability $\alpha_d=160.771$ a.u. for neutral Ca\cite{2014PhRvA..89b2506C} and $\alpha_d=162.4$ a.u. for neutral Na\cite{2015PhRvA..91a2705K}, respectively. All asymptotic forms of the excited states were adjusted to match the atomic energies in the separate atom limit given in Ref. \cite{NIST}.

Transition dipole moments for selected $\Sigma-\Sigma$ and $\Sigma-\Pi$ transitions are given in Figs. \ref{fig:dipolemoments} and \ref{fig:dipolemoments-pi}, and permanent dipole moments for the two lowest electronic states in Fig. \ref{fig:permdipolemoments}.

\subsection{Photoassociation}
\label{subsec:th_pa}

The single-photon photoassociation (PA) rate at temperature $T$ can be expressed as\cite{1987PhRvL..58.2420T,1994PhRvL..73.1352N,2006JPhB...39S.965J}
\begin{equation}
 K_{b}^{(1)}(T,I,\Delta) = \left\langle \frac{\pi v_{\mathrm{rel}}}{k^2} \sum_{\ell=0}^{\infty} (2\ell + 1) |S_b(\varepsilon, I, \Delta)|^2 \right\rangle ,
 \label{eq:PA1}
\end{equation}
where $\varepsilon = \hbar^2 k^2 / (2 \mu)$, $\mu$ is the reduced mass and $v_\mathrm{rel}$ is the relative velocity of the interacting particles, and $S_b$ is the scattering matrix element for photoassociating a molecular ion in the final bound ro-vibrational state $|b \rangle = |\ell', v'\rangle$ of the target electronic state. Averaging over relative velocities is implied by the $\left \langle \dots \right \rangle$. The PA laser field is represented by its intensity $I$, and the detuning from the target bound state $\Delta = E_\mathrm{b} - \hbar \omega$, where $E_\mathrm{b}$ is the energy of the target state.

At ultracold conditions the relative energy of interacting particles is such that their dynamics can be described well by the $s$-wave ($\ell=0$) scattering, or, in case of long-range nature of atom-ion interaction, the first few partial waves. 
By assuming a Maxwellian velocity distribution of a mixed-species ultracold gas, Eq. (\ref{eq:PA1}) can be written as an integral over collision energies\cite{1994PhRvL..73.1352N}
\begin{equation}
\begin{split}
 K_{b}^{(1)}(T,I,\Delta) = \frac{1}{h Q_T} \sum_{\ell=0}^{\infty} (2\ell + 1) 
 \int_{0}^{\infty} |S_b(\varepsilon, I, \Delta)|^2 e^{-\beta \varepsilon} d \varepsilon ,
 \label{eq:PA2}
\end{split}
\end{equation}
where $\beta = (k_B T)^{-1}$, $Q_T = (2 \pi \mu k_B T / h^2)^{3/2}$ is the translational partition function and $k_B$ is the Boltzmann constant.
For an isolated resonance, the scattering $S$ matrix can be approximated as\cite{1994PhRvL..73.1352N}
\begin{equation}
 |S_b(\varepsilon, I, \Delta)|^2 = \frac{\gamma_p \gamma_s(I,\varepsilon,\ell)}{(\varepsilon - \Delta)^2 + (\gamma / 2)^2} .
\end{equation}
Here, $\gamma = \gamma_p + \gamma_s(I,\varepsilon, \ell)$ is the total width (in units of energy) of the bound state, where $\gamma_p / \hbar$ is the spontaneous decay rate of the bound state resonance and $\gamma_s(I,\varepsilon, \ell) / \hbar$ is its stimulated emission rate. We assume that all other decay processes, such as molecular predissociation, are either negligible or not present.
For low laser intensities, the stimulated emission width to the bound state $|b \rangle$ can be expressed using Fermi's golden rule as
\begin{equation}
 \gamma_s(I,\varepsilon,v) = \frac{4 \pi^2 I}{c} |\langle b| d(R) |\varepsilon,\ell \rangle|^2 , 
 \label{eq:gammas}
\end{equation}
where the initial continuum state $| \varepsilon, \ell \rangle$ is energy-normalized and $d(R)$ is the transition dipole moment.
The scattering matrix element $S_b$ follows a general Lorenzian form that can be replaced by a delta function in the limit of small $\gamma$, which is satisfied for low laser intensities when $\gamma_p + \gamma_s \approx \gamma_p$. Consequently, for $\gamma_s / \gamma_p \ll 1$ and $\gamma_p / (2 \varepsilon) \ll 1$, we obtain 
\begin{equation}
 |S_b(\varepsilon, I, \Delta)|^2 = 2 \pi \gamma_s(I,\varepsilon,\ell) \delta(\varepsilon - \Delta) .
\end{equation} 
The Eq. (\ref{eq:PA2}) then takes a simple form
\begin{equation}
 K_{b}^{(1)}(T,I) = \frac{8 \pi^3}{h^2 c} \frac{I}{Q_T} e^{-1/2} |\langle b| d(R) |\varepsilon = k_B T/2, \ell \rangle|^2 ,
 \label{eq:PA3}
\end{equation}
where we assume the detuning $\Delta = k_B T / 2$, that yields maximal PA rate\cite{2006JPhB...39S.965J}. 
In an unpolarized sample 25\% of collisions between Na and Ca$^+$ happen along the singlet electronic state, analyzed in this study, while 75\% take place along the triplet potential curve. The rates presented in this study assume a spin-polarized sample and should be multiplied by an appropriate statistical weight if this is not the case.

\subsection{Spontaneous radiative decay}
\label{subsec:th_spont_decay}

Spontaneous emission or radiative decay processes can be described in terms of Einstein $A$ coefficients weighted by H\"onl-London factors. As illustrated in Fig. \ref{fig:scheme}, we are interested in the transition probabilities corresponding to spontaneous decay of the ro-vibrational level $b_Y=(v_Y,J_Y)$ in $\mathrm{E}^1\Sigma^+$ state ($Y=E$), as well as $\mathrm{B}^1\Sigma^+$ and $\mathrm{C}^1\Sigma^+$ states ($Y=B,C$), in case of two-photon excitation. Since the rotational states have much smaller energy splittings than the vibrational states and their main impact is on selection rules, we will restrict our analysis to rotational levels $J_E = 0$, $J_A = 1$, $J_{B,C} = 0,1,2$ and $J_X = 0$, and discuss them where appropriate.
The Einstein $A$ coefficients are given by
\begin{eqnarray}
 A_{v_E}^{\mathrm{tot}}(\mathrm{E}^1\Sigma^+) & = & A_{v_C}(\mathrm{C}^1\Sigma^+) 
    + A_{v_B}(\mathrm{B}^1\Sigma^+) + \nonumber \\ & & A_{v_A}(\mathrm{A}^1\Sigma^+) + A_{v_X}(\mathrm{X}^1\Sigma^+) ,
 \label{eq:AtotE}
\end{eqnarray}
and
\begin{eqnarray}
 A_{v_B}^{\mathrm{tot}}(\mathrm{B}^1\Sigma^+) & = & A_{v_A}(\mathrm{A}^1\Sigma^+) 
    + A_{v_X}(\mathrm{X}^1\Sigma^+), \\
 A_{v_C}^{\mathrm{tot}}(\mathrm{C}^1\Sigma^+) & = & A_{v_A}(\mathrm{A}^1\Sigma^+) 
    + A_{v_B}(\mathrm{B}^1\Sigma^+) + \nonumber \\ & &  A_{v_X}(\mathrm{X}^1\Sigma^+) .
 \label{eq:AtotB}
\end{eqnarray}
Here, all other allowed transitions, including ro-vibrational transitions within the electronic states, spontaneous emission from $\mathrm{E}^1\Sigma^+ \rightarrow \mathrm{D}^1\Sigma^+$, and transitions from $^{1}\Sigma$ into $^{1}\Pi$ states, have smaller probabilities and can be neglected.
The individual contributions $A_{v_Y}(\mathrm{Y})$ correspond to the sum of allowed $R$, $Q$, and $P$-branch\cite{Herzberg} transitions into the electronic state $\mathrm{Y}'$, where each branch $\alpha=R,Q,P$ includes both bound-bound and bound-continuum transitions: 
\begin{equation}
 A_{v_Y}^{\alpha}(\mathrm{Y}) = \sum_{v_{Y'}} A_{v_Y v_{Y'}}^{\alpha}(\mathrm{Y'}) 
    + \int A_{v_Y}^{\alpha} (\mathrm{Y},\varepsilon') \mathrm{d}\varepsilon' .
 \label{eq:AY1}
\end{equation}
Here, the transitions $A_{v_Y v_{Y'}}^{\alpha}$, from the initial state $|v_Y,J_Y \rangle$ to $|v_{Y'},J_{Y'} \rangle$, are given by 
\begin{eqnarray}
  A_{v_Y v_{Y'}}^\alpha(\mathrm{Y}) & = & \frac{4 e^2 (\omega_{v_Y v_{Y'}}^\alpha)^3}{3 \hbar c^3}  W_{J_{Y'}}^{\alpha} |\langle v_{Y'} J_{Y'} | D_{Y' Y} | v_Y J_Y \rangle|^2 \nonumber \\
  A_{v_Y}^\alpha(\mathrm{Y,\varepsilon}) & = & \frac{4 e^2 \omega_{v_Y}^\alpha(\varepsilon)^3}{3 \hbar c^3} W_{J_{Y'}}^{\alpha} |\langle \varepsilon \ell | D_{Y' Y} | v_Y J_Y \rangle|^2 .
 \label{eq:AY2}
\end{eqnarray}
Here, $\hbar \omega_{v_Y v_{Y'}}^{\alpha} = |E_{v_Y J_Y} - E_{v_{Y'} J_{Y'}}|$ and $\hbar \omega_{v_Y}^{\alpha}(\varepsilon) = |E_{\varepsilon \ell} - E_{v_Y J_Y}|$ are the frequencies for the bound-bound and bound-free transition, respectively, while $D_{Y'Y}(R)$ is the dipole transition moment between the initial and final electronic states Y and $\mathrm{Y'}$ (\textit{e.g.}, $D_{AE}(R)$ for the transition from Y$=\mathrm{E}^1\Sigma^+$ to $\mathrm{Y'}=\mathrm{A}^1\Sigma^+$), and $W_{J_{Y'}}^{\alpha}$ are the H\"onl-London factors for the branch $\alpha$\cite{Herzberg}. Note that the most significant contribution from the continuum at ultracold temperatures will occur for small values of $\varepsilon$, approving the use of the dipole transition moment in free-bound transitions. 

The lifetime of the level $(v_Y, J_Y)$ is given by 
\begin{equation}
 \tau_{v_Y} = \frac{1}{A_{v_Y}^{\mathrm{tot}}(Y)} \,.
\end{equation}
We also define the state-to-state branching ratio for the spontaneous radiative emission involving the initial and final states $\mathrm{Y} (v_Y,J_Y)$ and $\mathrm{Y'} (v_{Y'},J_{Y'})$, respectively, as 
\begin{equation}
 r_{v_{Y'} J_{Y'}}^{v_Y J_Y}(\alpha) = \frac{A_{v_Y v_{Y'}}^\alpha(\mathrm{Y'})} {A_{v_Y}^{\mathrm{tot}}(Y)} \, .
\end{equation}

\section{Results}
\label{sec:results}

\subsection{Dipole transition matrix elements}
\label{subsec:res_transitions}

\begin{figure*}[t]
 \centering
 \includegraphics[clip,width=\linewidth]{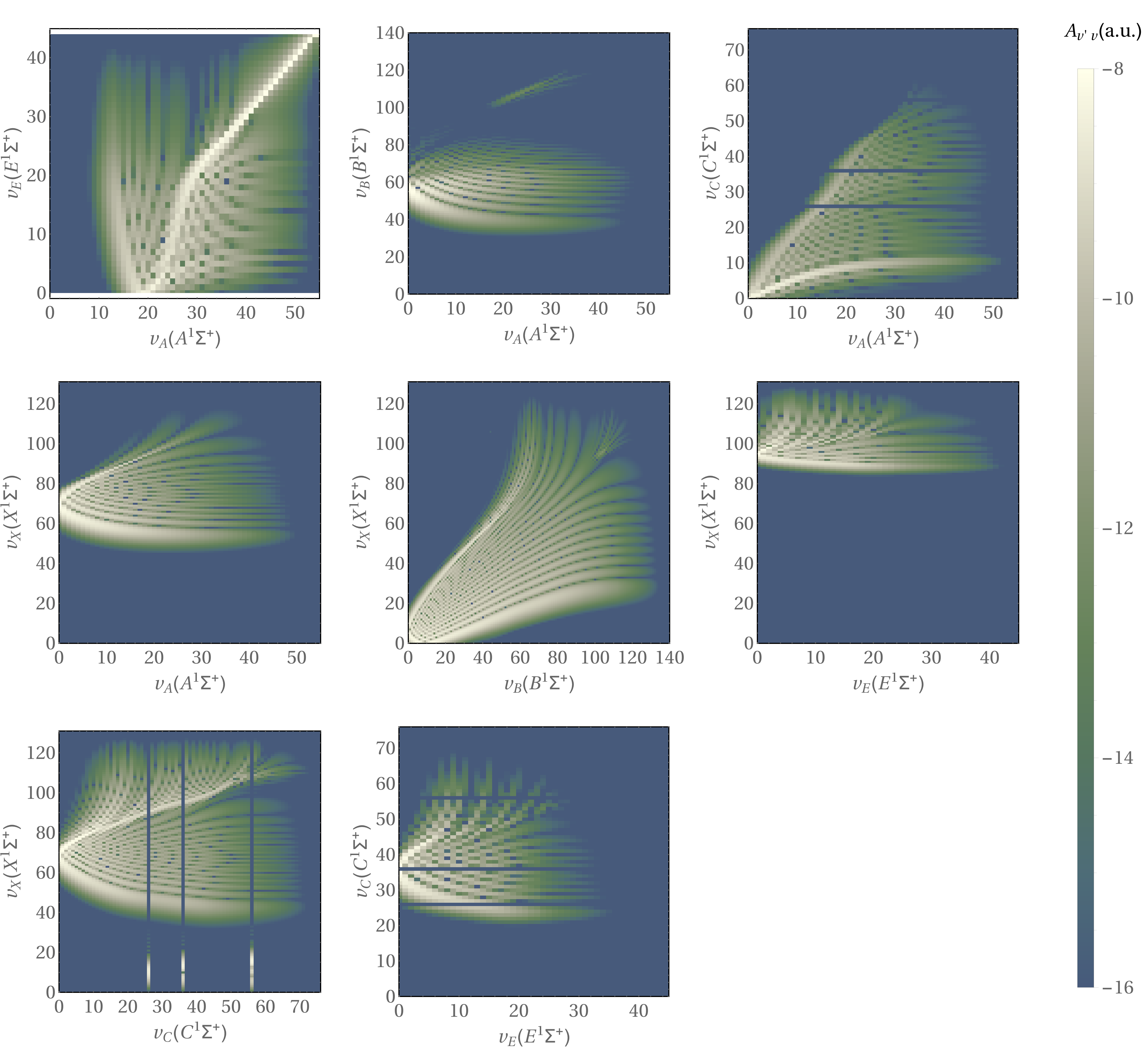}
 \caption{(Color online) Einstein $A$ coefficients (in atomic units) for bound-bound dipole transitions for selected pairs of electronic states: 
 $\mathrm{E}^1\Sigma^+ \leftrightarrow \mathrm{A}^1\Sigma^+$ (top left),  
 $\mathrm{B}^1\Sigma^+ \leftrightarrow \mathrm{A}^1\Sigma^+$ (top middle),  
 $\mathrm{C}^1\Sigma^+ \leftrightarrow \mathrm{A}^1\Sigma^+$ (top right), 
 $\mathrm{X}^1\Sigma^+ \leftrightarrow \mathrm{A}^1\Sigma^+$ (center left), 
 $\mathrm{X}^1\Sigma^+ \leftrightarrow \mathrm{B}^1\Sigma^+$ (center), 
 $\mathrm{X}^1\Sigma^+ \leftrightarrow \mathrm{E}^1\Sigma^+$ (center right). 
 $\mathrm{X}^1\Sigma^+ \leftrightarrow \mathrm{C}^1\Sigma^+$ (bottom left). 
 $\mathrm{C}^1\Sigma^+ \leftrightarrow \mathrm{E}^1\Sigma^+$ (bottom middle).
 }
 \label{fig:Acoefs}
\end{figure*}

To determine the optimal pathways for production of molecular ions in lower-energy electronic states, we calculated dipole transition matrix elements and Einstein $A$ coefficients between bound vibrational levels for all pairs of electronic states energetically below $E^1\Sigma^+$ asymptote and allowed by symmetry. 
The matrix elements, given in Eqs. (\ref{eq:AY1}) and (\ref{eq:AY2}), were evaluated numerically by diagonalizing the radial Schr\"odinger equation using mapped Fourier grid method (MFGR)\cite{1999JChPh.110.9865K} to simultaneously obtain bound and quasi-discretized continuum spectrum. The MFGR calculation was performed with no couplings between different potential curves, and assuming a variable grid step size determined on the total box size ($R_\mathrm{max}=5 \times 10^4$ $a_0$) and mapping potential determined from the local momentum, while the accuracy of the wave functions in the highly oscillatory short-range region was ensured by small scaling factor $\beta$, resulting in at least 20 points per a single oscillation of the wave function. 
The continuum wave functions were found to be in excellent agreement with a calculation performed using renormalized Numerov method\cite{ren_numerov} for continuum energies larger than about 500 nK, below which Fourier grid method requires a larger box size.

The Einstein $A$ coefficients sufficiently large to be relevant for the analysis of the possible transitions pathways are shown in Fig. \ref{fig:Acoefs}. 
Specifically, these include the transitions involving the $E^1\Sigma^+$ state $\left(\mathrm{to}\hspace{0.3em} A^1\Sigma^+ , X^1\Sigma^+, C^1\Sigma^+ \right)$, $A^1\Sigma^+$ state $\left(\mathrm{to}\hspace{0.3em} X^1\Sigma^+ , B^1\Sigma^+ , C^1\Sigma^+ , E^1\Sigma^+ \right)$, and the dipole transitions relevant for 2-photon exitation scheme $\left( B^1\Sigma^+ \leftrightarrow X^1\Sigma^+ , C^1\Sigma^+ \leftrightarrow X^1\Sigma^+ \right)$. 
The calculated coefficients play a key role in selection of electronic states and analyzing possible single- or two-photon optical pathways for formation of ground state molecular ions (see Fig. \ref{fig:scheme}).
For example, from the Fig. \ref{fig:Acoefs}(top left) it is evident that the molecular ions produced in the last few bound ro-vibrational levels of the $E^1\Sigma^+$ state, such as $b_\mathrm{max}=(J'=0,v'=43)$, strongly favor a dipole transition into the highest (near dissociation) vibrational levels of the $A^1\Sigma^+$ electronic state, with a very small fraction of population being able to transition directly into the ground $X^1\Sigma^+$ state. 
The features present as three ``lines'' in the transitions involving C$^1\Sigma^+$ potential are caused by tunneling through the shallow potential well in the short-range region (see Fig. \ref{fig:potentials_sigma}).

\subsection{Single photon excitation}
\label{subsec:res_pa}

\subsubsection{Photoassociation and spontaneous relaxation}
A possible approach to formation of the molecular ions in the ground electronic state, analyzed in this study, involves a single photon photoassociation (PA) of the NaCa$^+$ molecular ions in the excited $\mathrm{E}^1\Sigma^+$ state from a pair of Na($^2$S) atoms and Ca$^+$($^2$S) ions colliding at ultralow energy via the $\mathrm{A}^1\Sigma^+$ PEC, followed by a spontaneous relaxation into the $\mathrm{X}^1\Sigma^+$ (see Fig. \ref{fig:scheme}). 
The choice of the initial state is motivated by ongoing experiments\cite{2014ApPhB.114...75S}, the fact that a magneto-optical trap (MOT) of Na atoms is readily available, and the analysis of dipole transitions given in the previous section. 

\begin{figure}[t]
 \centering
 \includegraphics[clip,width=\linewidth]{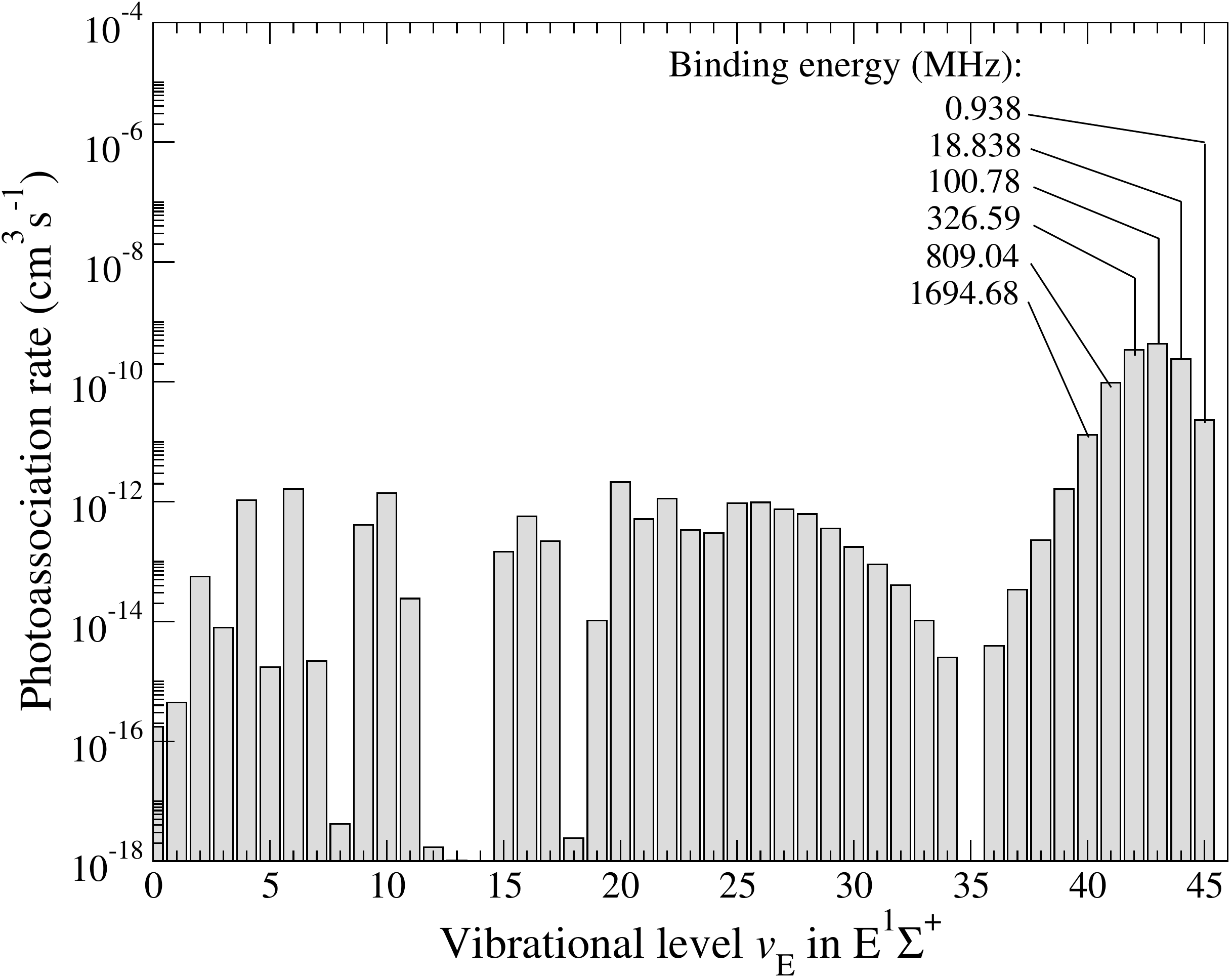}
 \caption{(Color online) Photoassociation rate of produced NaCa$^+$ in the $\mathrm{E}^1\Sigma^+$ state in dependence on the target ro-vibrational level $(v_E,J_E=0)$. Binding energies of the highest levels are also indicated.}
 \label{fig:KPA}
\end{figure}

The initial scattering wave function, required to compute the stimulated emission width and photoassociation rate (Eqs. (\ref{eq:PA2} \& \ref{eq:gammas})), was calculated using renormalized Numerov method\cite{ren_numerov} for approximately one hundred collision energies on a nonuniform grid between $10^{-15}$ and $10^{-8}$ Hartree, for internuclear separations up to $5\times 10^4$ $a_0$. The PA rate of NaCa$^+$ molecular ions in the $\mathrm{E}^1\Sigma^+$ state, calculated according to Eqs. (\ref{eq:PA2} \& \ref{eq:PA3}), is given in Fig. \ref{fig:KPA} as a function of the final vibrational level for $J'=0$. 
Here, we assumed a Maxwell-Boltzmann distribution of atom-ion collision energies at the average gas temperature of $T = 3.16$ $\mu\mathrm{K}$, the PA laser intensity of 1 kW/cm$^2$ at the optimal wavelengths (detuned by $\Delta_v = k_B T/2$ from the resonant transition frequency for the vibrational state $v$), and the reduced mass $\mu(\mathrm{^{23}Na^{40}Ca}) = 26612.439053459$ $\mathrm{amu}$.

\begin{figure}[t]
 \centering
 \includegraphics[clip,width=\linewidth]{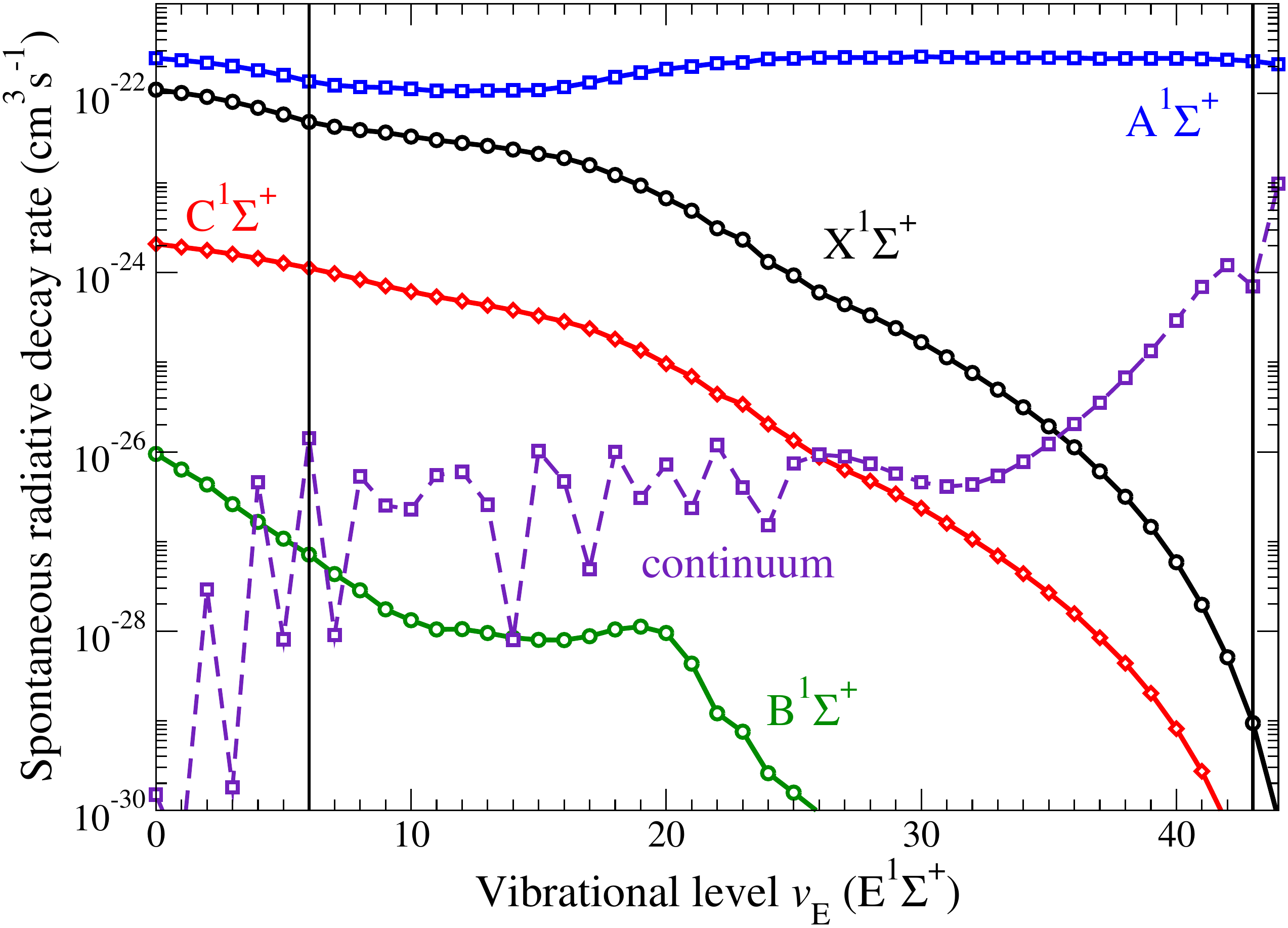}
 \caption{(Color online) Total (sum over all target ro-vibrational states) spontaneous radiative decay rates from the initial state $\mathrm{E}^1\Sigma^{+}(v_E, J_E=0)$ state of NaCa$^+$ into lower-energy electronic states. The vibrational levels of interest for cold molecular ion formation are indicated by vertical lines.}
 \label{fig:Edecay}
\end{figure}

At the considered experimental conditions, the dominant decay process for the excited electronic states of molecular ions will spontaneous radiative decay to lower states and the continuum. The calculated radiative decay rates for the $E^1\Sigma^+$ state, of interest in our study, are given in Fig. \ref{fig:Edecay}. 
The spontaneous decay to the continuum for the lower-energy electronic states was included as converged sums over the quasi-discretized continuum (all discretized continuum states that contribute more than 0.01 \% to the total sum were included; typically the sums included between 300 and 800 states).

\subsubsection{Formation rates in $E^1\Sigma^+$ and $X^1\Sigma^+$ states}
\label{subsec:formation_ground_state}

\begin{figure}[t]
 \centering
 \includegraphics[clip,width=\linewidth]{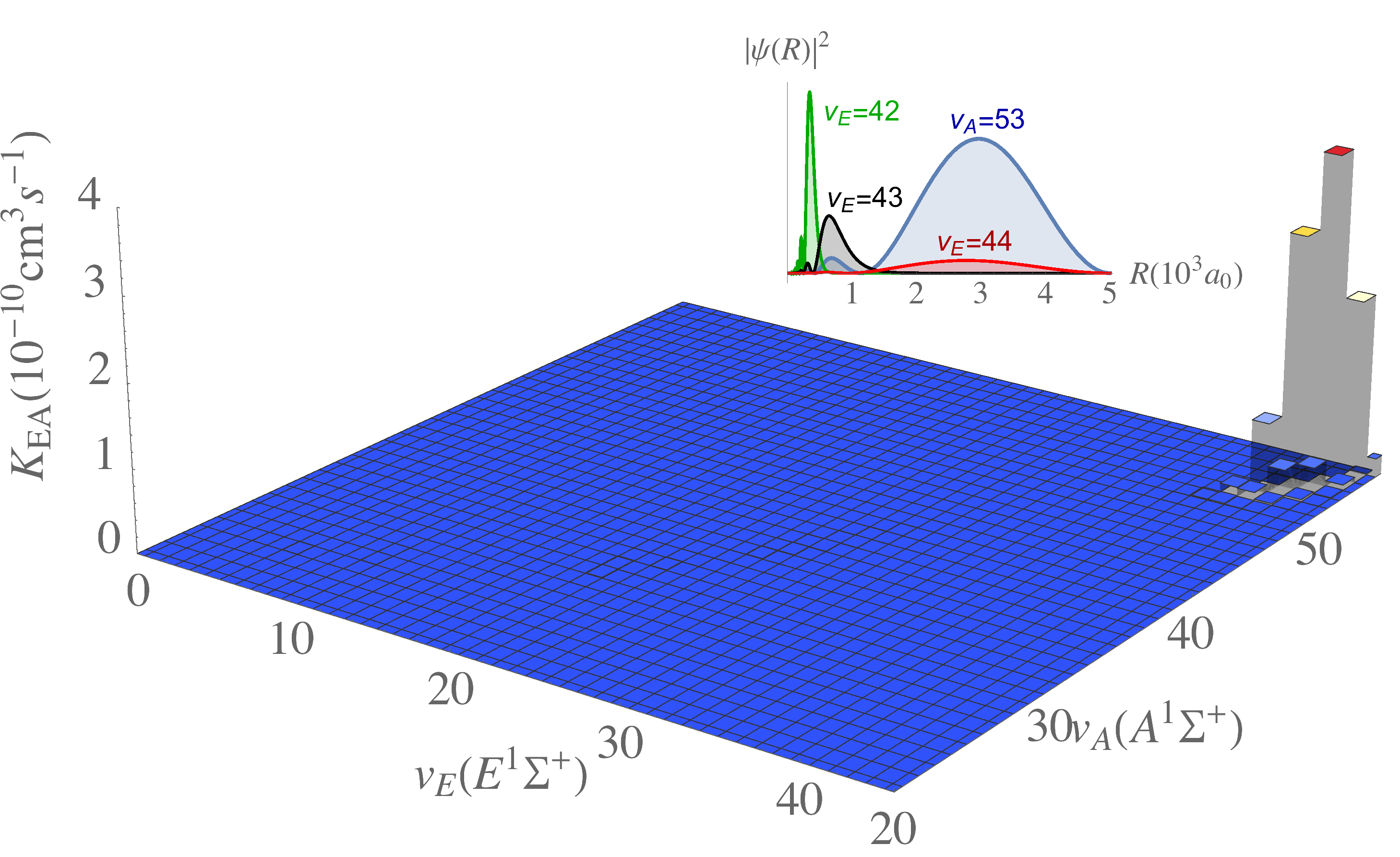}
 \includegraphics[clip,width=\linewidth]{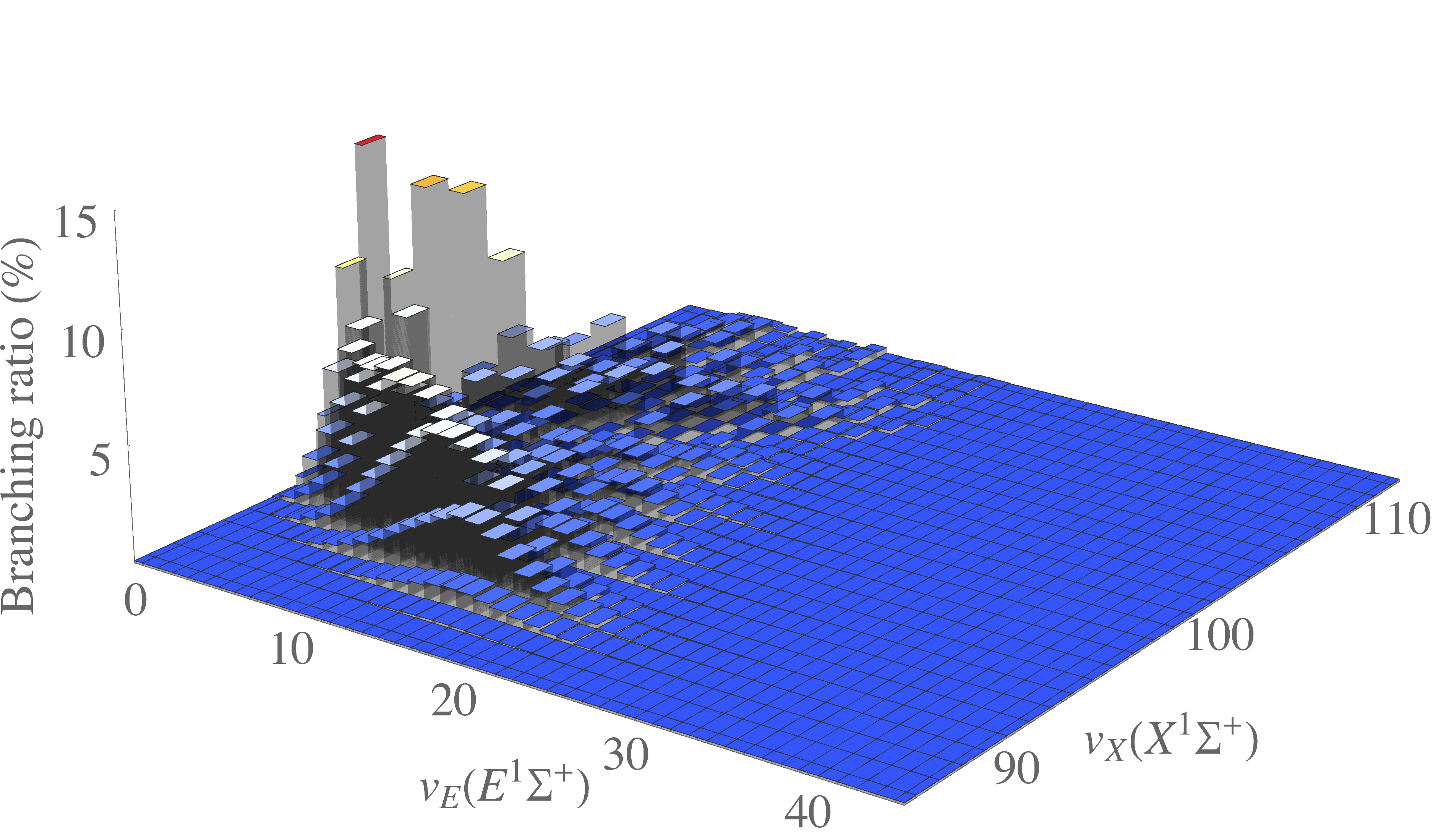}
 \includegraphics[clip,width=\linewidth]{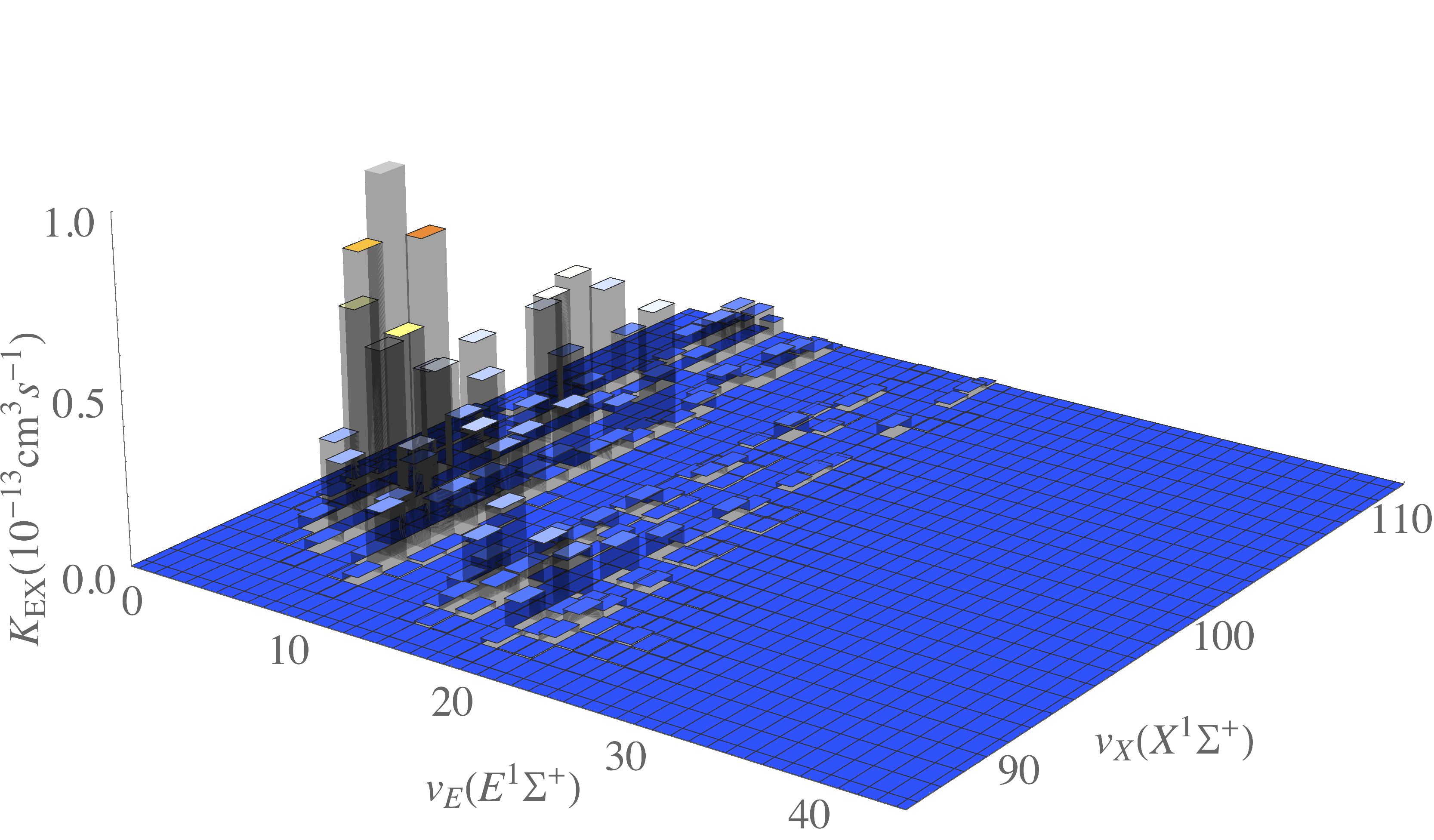}
 \caption{(Color online) \textit{Top:} Formation rate $K_{\mathrm{EA}}(v_E,v_A) = K^{PA}_{v_E,J_E=0} \; r_{v_A,J_A=1}^{v_E,J_E=0}$ of NaCa$^+$ in the $\mathrm{A}^1\Sigma^+$ state, by spontaneous decay from the $\mathrm{E}^1\Sigma^+$ state, as a function of vibrational levels.
 \textit{Middle:} Branching ratio $r_{v_X,J_X=1}^{v_E,J_E=0}$ for spontaneous radiative decay into the electronic ground state of CaNa$^{+}$ as a function of vibrational levels of $\mathrm{E}^1\Sigma^+$ and $\mathrm{X}^1\Sigma^+$ states.
 \textit{Bottom:} Formation rate of CaNa$^{+}$ in the $\mathrm{X}^1\Sigma^+$ state by spontaneous radiative decay of the photoassociated NaCa$^{+}$ in the $\mathrm{E}^1\Sigma^+$ state as a function of vibrational levels.}
 \label{fig:KAE_3d}
\end{figure}

The fact that the long-range part of atom-ion interaction potentials is dominated by the $R^{-4}$ term, implies that the PA will favor the most extended bound states to a greater extent than in case of ultracold neutral atoms. This is consistent with our results (Fig. \ref{fig:KPA}), where we obtained the most significant PA rates for the last five vibrational levels, $v_E = 40-45$. For the physical parameters listed above, the maximal PA rate, $K^\mathrm{PA}_\mathrm{max} = 4.3 \times 10^{-10}$ cm$^3$s$^{-1}$, suggesting that thousands of molecular ions per second could be formed in the ro-vibrational level $b_{E}^\mathrm{max}=(v_E=43,J_E=0)$ of $\mathrm{E}^1\Sigma^+$ state in high-density trapped samples. Here, the stimulated emission induced by the PA laser is neglected.

The calculated Einstein $A$ coefficients (Fig. \ref{fig:Acoefs}) and dipole transition spontaneous emission rates (Fig. \ref{fig:Edecay}) indicate that majority of the molecular ions photoassociated in $b_\mathrm{max}$ state of $E^1\Sigma^+$ will spontaneously decay into high vibrational levels of the $A^1\Sigma^+$ state and form loosely bound molecular ions (Fig. \ref{fig:KAE_3d}, \textit{top}), while a small fraction will be lost to the continuum. 
The branching ratios $b^{v_X J_X}_{v_E J_E}(\alpha)$, for the spontaneous emission from $E^1\Sigma^+$ into $X^1\Sigma^+$ state (Fig. \ref{fig:KAE_3d}, \textit{middle}), indicate that significant formation of the ground state molecular ions via this process can occur only from deeply bound vibrational levels $(v_E<20)$ of $E^1\Sigma^+$ state ($10-15$\% for $v_E=0-5$, about 4\% for $v_E=20$), while high vibrational levels, accessible by the PA, have insignificant branching ratios.

Formation rates for (NaCa)$^+$ in the ground electronic state by spontaneous radiative decay of the photoassociated ions in the $E^1\Sigma^+$ state are given in (Fig. \ref{fig:KAE_3d}, \textit{bottom}). The maximum formation rate, $K^\mathrm{PA}_\mathrm{v_E,v_X} = 1.05 \times 10^{-13}$ cm$^3$s$^{-1}$, is obtained for $(v_E=6,v_X=92)$, where the overlap of the inner turning point in $E^1\Sigma^+$ state and outer turning point of the ground state is optimal. 
The highest total formation rate in the ground state, $K^\mathrm{PA}_\mathrm{v_E=6} =\sum_{v_X} K^\mathrm{PA}_\mathrm{v_E=6,v_X} = 4.9 \times 10^{-13}$ cm$^3$s$^{-1}$. 

\subsection{Two-photon excitation via an intermediate state}

\begin{figure}[t]
 \centering
 \includegraphics[clip,width=0.9\linewidth]{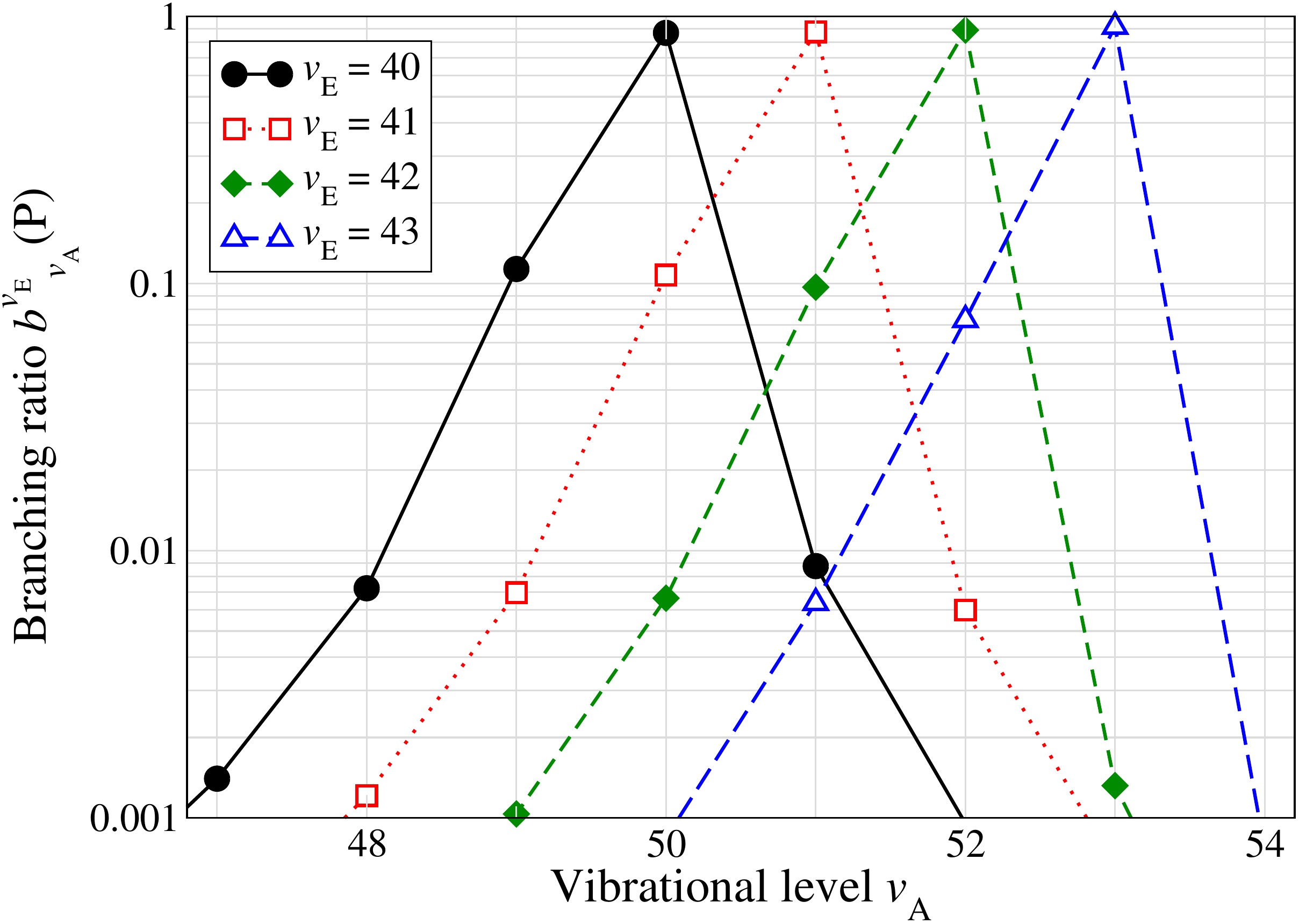}
 \caption{(Color online) Branching ratios $b^{v_E J_E=0}_{v_A J_A=1}(\alpha)$ for spontaneous emission from $\mathrm{E}^1\Sigma^+$ to $\mathrm{A}^1\Sigma^+$ state.}
 \label{fig:bratio_EtoA}
\end{figure}

In the previous section, we have shown that the ground state formation rates are limited by the fact that high ro-vibrational levels of the $E^1\Sigma^+$ state, accessible to PA, will almost exclusively decay to high ro-vibrational levels near the dissociation limit of the $A^1\Sigma^+$ state (\ref{fig:bratio_EtoA}). These, spatially very extended, states have very long lifetimes with respect to vibrational and rotational transitions, and take minutes to spontaneously decay into the ground electronic state, resulting in very small production rates.

The proposed formation scheme could be improved by introducing another excitation step, where the second laser would be used to transfer the population from the $A^1\Sigma^+$ to an intermediate state which has more favorable Franck-Condon factors with the electronic ground state (Fig. \ref{fig:scheme}). The entire process can be considered in two separate steps: i) the PA of molecular ions in the $E^1\Sigma^+$ state, followed by their spontaneous decay to the $A^1\Sigma^+$ state; and ii) the stimulated bound-bound transition $A^1\Sigma^+ \rightarrow B^1\Sigma^+$, followed by the spontaneous decay into the ground state.

The optimal intermediate state can be selected based on dipole transition matrix elements (Fig. \ref{fig:Acoefs}). 
We performed the analysis for two possible intermediate electronic states: $B^1\Sigma^+$ and $C^1\Sigma^+$. These states have favorable dipole transition matrix elements for the transition to the $X^1\Sigma^+$ state, while remaining accessible from the highly excited ro-vibrational levels of $A^1\Sigma^+$ state.

\begin{figure}[t]
 \centering
 \includegraphics[clip,width=\linewidth]{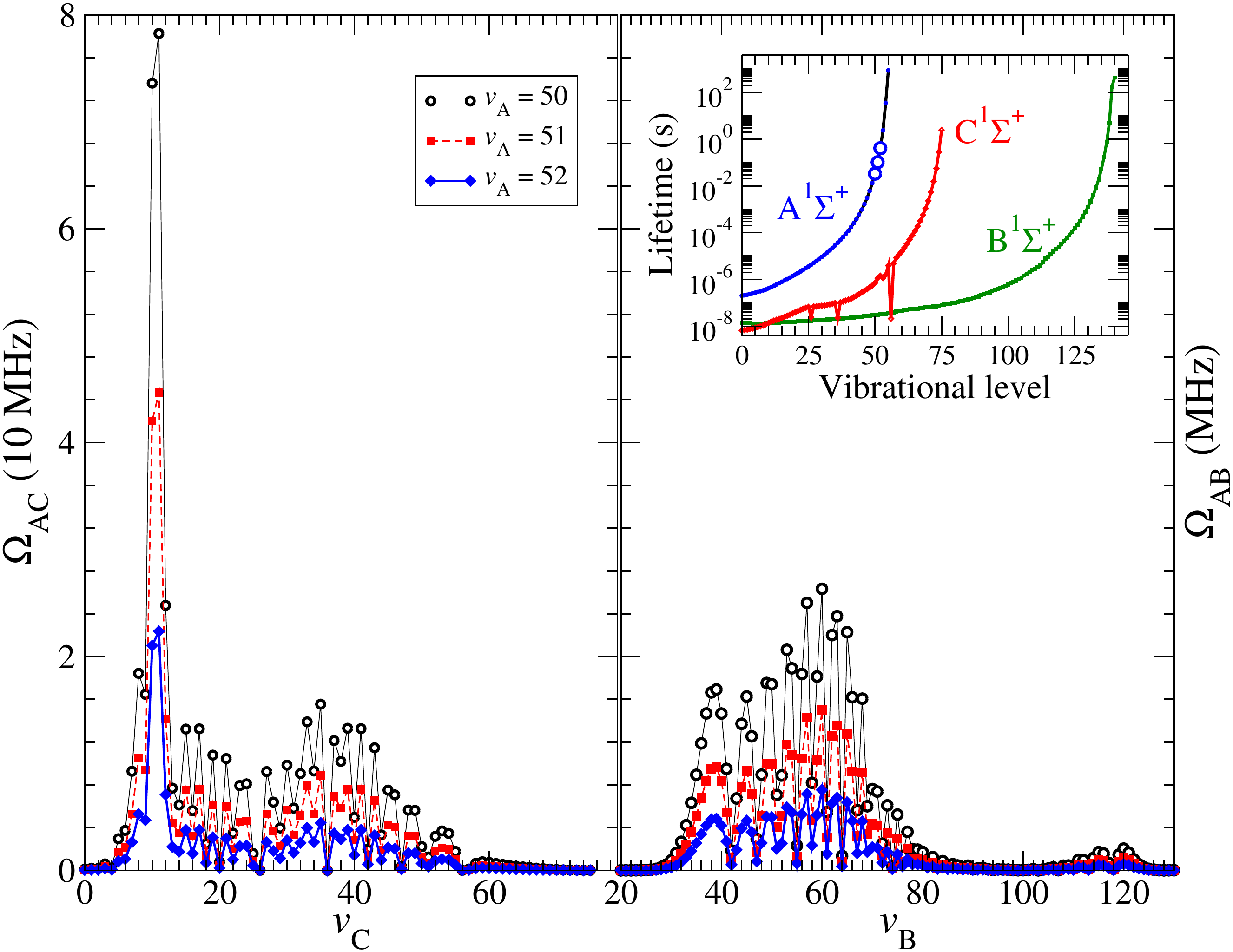}
 \caption{(Color online) Rabi frequencies for stimulated transitions $\mathrm{A}^1\Sigma^+ \leftrightarrow \mathrm{C}^1\Sigma^+$ (left panel) and $\mathrm{A}^1\Sigma^+ \leftrightarrow \mathrm{B}^1\Sigma^+$ (right panel) in NaCa$^+$ for initial ro-vibrational levels $(v_A=50\dots52, J_A=1)$. \textit{Inset:} Lifetimes of ro-vibrational levels $(v_Y,J_Y=0)$ for $Y=A,B,C$. The considered initial ro-vibrational levels in $\mathrm{A}^1\Sigma^+$ are indicated by circles.}
 \label{fig:RabiABC}
\end{figure}

To estimate the efficiency of the second excitation-relaxation step we calculated Rabi frequencies $\Omega_{AB}$ and $\Omega_{AC}$ for the transitions in the two-level system approximation\cite{demtroder2007laserspektroskopie}
\begin{equation}
 \Omega_{AY} = \frac{e}{\hbar} \sqrt{\frac{2 I_2}{\varepsilon_0 c}} |\langle v_A J_A| D_{AY} |v_Y J_Y \rangle| ,
\end{equation}
where $Y=\{B,C\}$, $I_2$ is the intensity of a monochromatic cw laser tuned to the transition frequency (no detuning), and $c$ is the speed of light in vacuum, $e$ is the electron charge, and $\varepsilon_0$ is the vacuum permittivity. 
Fig. \ref{fig:RabiABC} shows the Rabi frequencies for the initial ro-vibrational states $(v_A, J_A) = (50,0), (51,0), (52,0)$, predicted to be significantly populated after the first excitation-relaxation step. The intensity of the second laser was taken to be $I_2 = 1000$ W/cm$^{2}$.
The spontaneous radiative decay rates, calculated from Eqs. (\ref{eq:AtotE}-\ref{eq:AtotB}) and given in the Fig. \ref{fig:RabiABC} (inset), indicate that the spontaneous emission from either intermediate state to the ground electronic state will be two or more orders of magnitude faster than the decay back into $A^1\Sigma^+$ state. Therefore, the losses of the molecular ions due to the decay to different ro-vibrational levels or the continuum will be small and are ignored in the rest of the analysis.

\begin{figure}[t]
 \centering
 \includegraphics[clip,width=\linewidth]{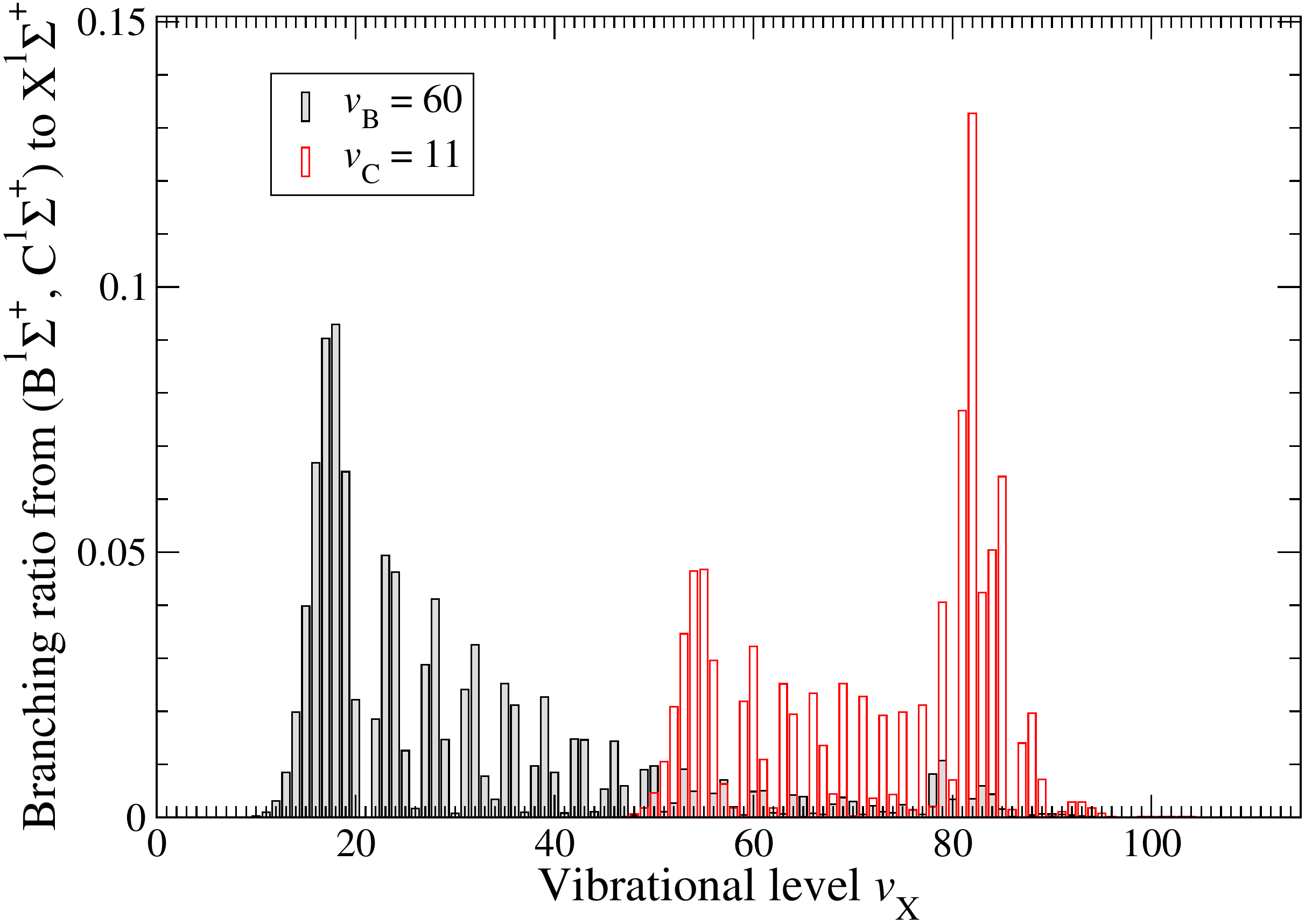}
 \caption{(Color online) Branching ratios for spontaneous emission from $\mathrm{B}^1\Sigma^+(v_B=60,J_c=0)$ and $\mathrm{C}^1\Sigma^+(v_c=11,J_c=0)$ to $\mathrm{X}^1\Sigma^+(v_X, J_X=0)$.} 
 \label{fig:rovibX-2photon}
\end{figure}

The branching ratios $b^{v_B J_B}_{v_X J_X}(\alpha)$ and $b^{v_C J_C}_{v_X J_X}(\alpha)$, for the spontaneous emission from $\mathrm{B}^1\Sigma^+(v_B=60,J_B=0)$ and $\mathrm{C}^1\Sigma^+(v_C=11,J_C=0)$ into the ro-vibrational levels $(v_X, J_X=0)$ of $\mathrm{X}^1\Sigma^+$ state, are given in Fig. \ref{fig:rovibX-2photon}. The illustrated ro-vibrational levels have the highest Rabi frequencies for the stimulated transition from the $\mathrm{A}^1\Sigma^+$ state, and, consequently, require the lowest laser intensity $I_2$ to saturate the transitions. The spontaneous emission from $\mathrm{B}^1\Sigma^+$ state will result in formation of molecular ions mainly in deeply bound vibrational levels, between about $v_B=10$ and $v_B=60$. The population distribution formed by the spontaneous emission from the $\mathrm{C}^1\Sigma^+$ state will be centered at higher vibrational levels, mostly between $v_B=50$ and $v_B=90$, peaking around $v_B=82$. 

\begin{table}
\caption{Formation rates $K_X$ for molecular ions produced in $\mathrm{X}^1\Sigma^+$ state via an electronic state $Y$ for optimal production pathways. The optimal single-photon formation rate ($Y=E$) is given in the last row.}
\begin{ruledtabular}
\begin{tabular}{ccccc}
$(v_E, v_A)$ & $v_B$ & $v_C$ & $\Omega_{AY}$(MHz) & $K_\mathrm{X}$(cm$^3$/s) \\
\hline
(43,53) & 60 &    & 0.313 & 3.83 $\times 10^{-10}$ \\
        &    & 11 & 9.314 & 3.73 $\times 10^{-10}$  \\
(42,52) & 60 &    & 0.751 & 2.95 $\times 10^{-10}$  \\
        &    & 11 & 22.351 & 2.87 $\times 10^{-10}$  \\
(41,51) & 60 &    & 1.502 & 8.17 $\times 10^{-11}$  \\
        &    & 11 & 44.679 & 7.97 $\times 10^{-11}$  \\ \hline

(6, *)   &    &    &       & 4.90 $\times 10^{-13}$
\end{tabular}
\end{ruledtabular}
\label{table1}
\end{table}

Predicted formation rates for ground state molecular ions are given in Table \ref{table1}. The required wavelengths of listed optical excitations are approximately $\lambda_{1}^{\mathrm{PA}} = 589.457$ nm, $\lambda_{AB} = 827.532$ nm, and $\lambda_{\mathrm{AC}} = 751.873$ nm.

\subsection{Vibrational relaxation of the population in the ground electronic state}

\begin{figure}[t]
 \centering
 \includegraphics[clip,width=\linewidth]{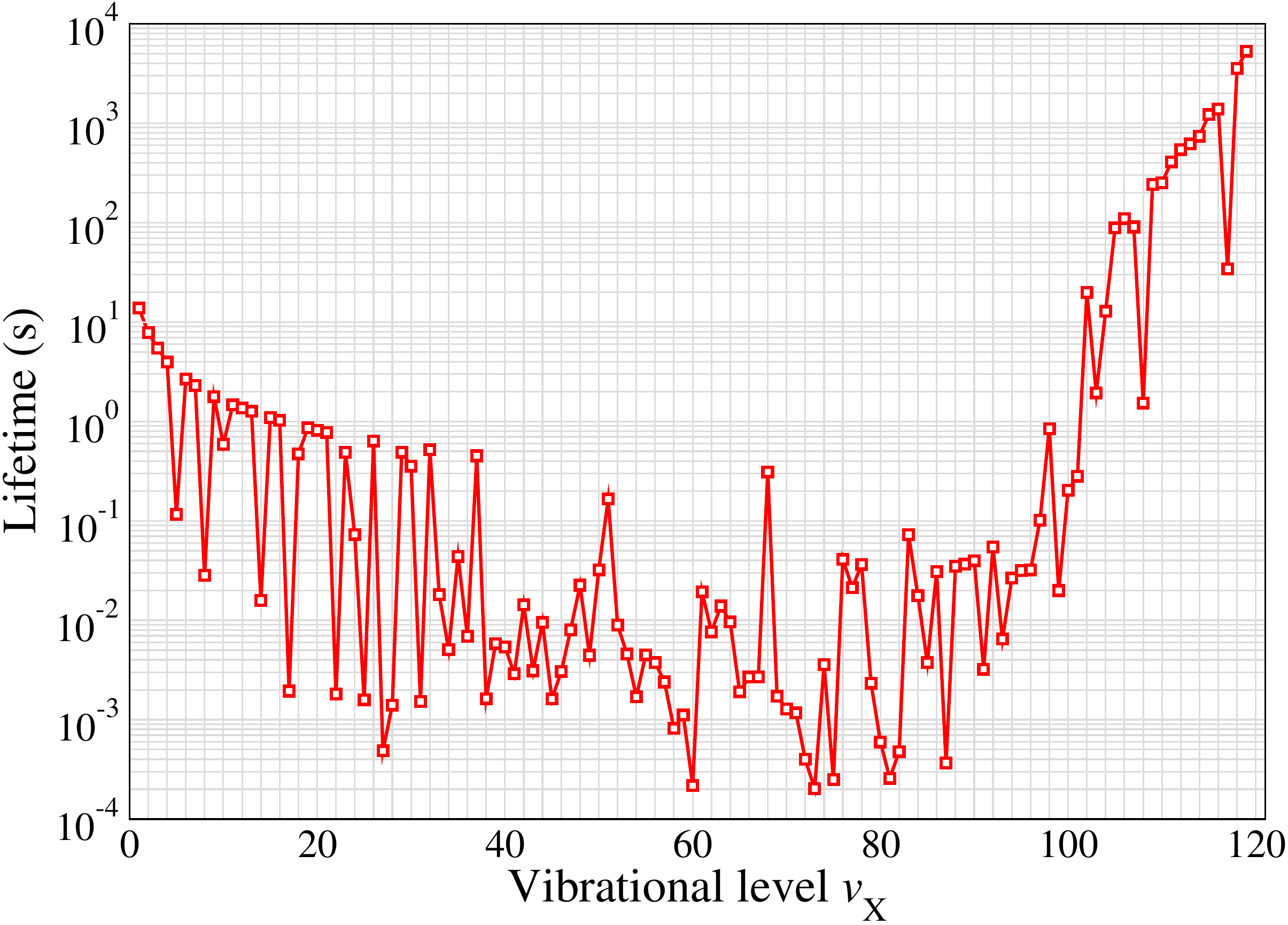}
 \caption{(Color online) Lifetime of vibrational levels $v_X$ in $\mathrm{X}^1\Sigma^+$.} 
 \label{fig:lifetimeXtoX}
\end{figure}

Once a molecular ion is formed in an excited vibrational level $v_X$, it will decay in a cascade of spontaneous emission steps into lower vibrational levels until it reaches $v_X=0$. The radiative cascade will ultimately produce a distribution of rotational states determined by the selection rule $\Delta J = \pm1$ imposed on the transitions.
To determine the relaxation time, we calculated lifetimes of the vibrational levels $v_X$ as sums of all possible decay pathways (Fig. \ref{fig:lifetimeXtoX}). Since the lifetimes of the lowest vibrational levels, on the order of 1-10 seconds, are much larger than the lifetimes of the intermediate vibrational levels, below $v_X<90$, populated in suggested two-photon formation pathways, the total cascade time $\tau_\mathrm{cascade}$ can be estimated as a sum of the lifetimes of the final three transitions, resulting in $\tau_\mathrm{cascade}\approx26$ seconds.

\begin{figure}[t]
 \centering
 \includegraphics[clip,width=0.93\linewidth]{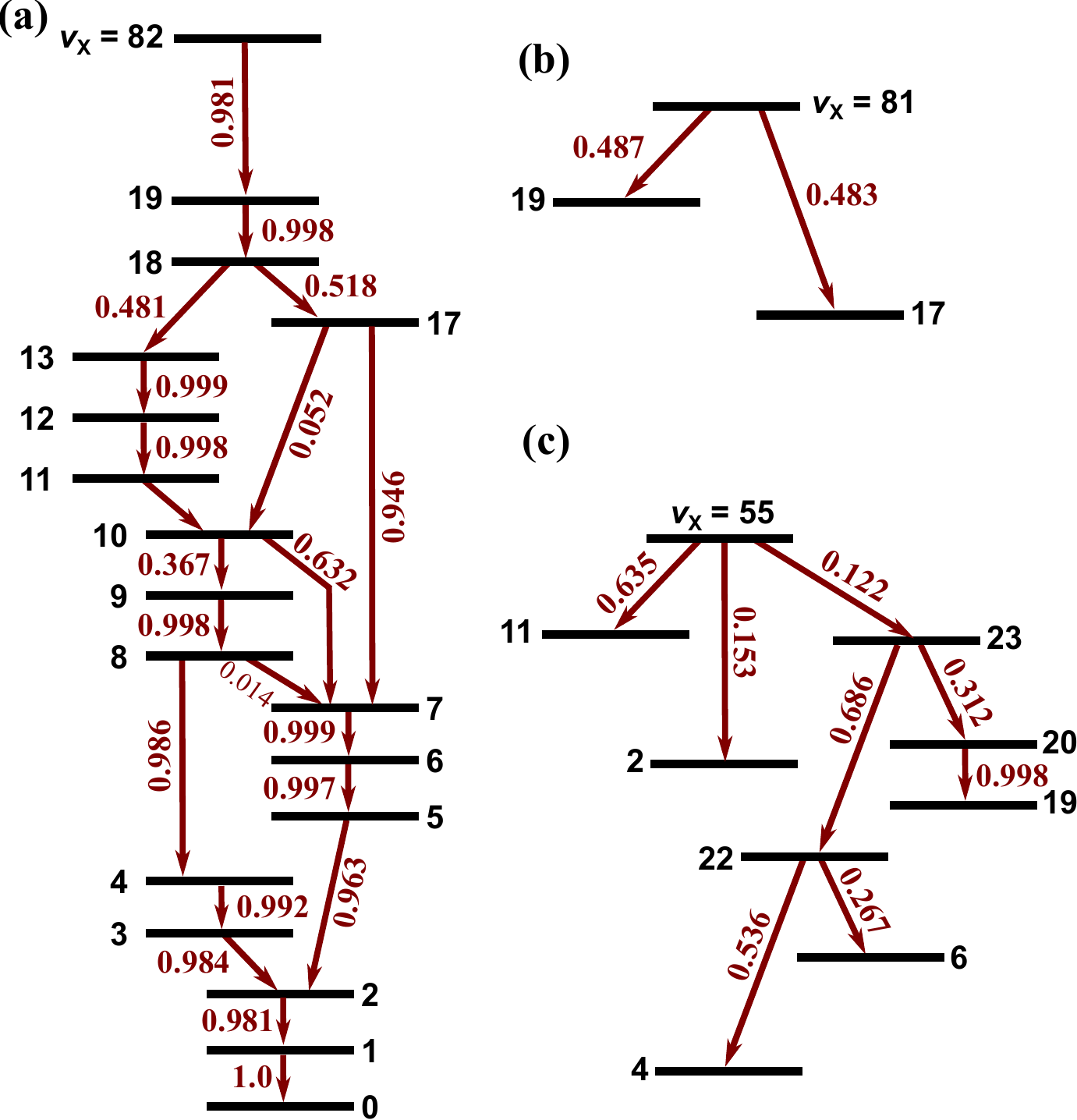}
 \caption{(Color online)  Radiative cascade in $\mathrm{X}^1\Sigma^+$ state initiated from vibrational levels $v_X=82$ (a), $v_X=81$ (b), and $v_X=55$ (c). Transition probabilities are given on arrows. } 
 \label{fig:cascade}
\end{figure}

A better estimate can be made by analyzing most probable decay pathways. We took the levels $v_X=55,81,82$ and $v_X=18-20$ to be representative of the decays from $\mathrm{C}^1\Sigma^+$ and $\mathrm{B}^1\Sigma^+$ states, respectively (see Fig. \ref{fig:rovibX-2photon}). Vibrational radiative cascades originating from these levels are illustrated in Fig. \ref{fig:cascade}. The cascades from $v_X=82$ and $v_X=55$ are of particular interest, since other initial states either revert to them after a single transition or are already included as intermediate steps.
The cascade times from $v_X=82$, depending on the branch, are: 28.54 s (via $v_X=17$), 40.52 s (via $v_X=13,9$), and 33.63 s (via $v_X=13,7$). The first transition, $v_X=82 \rightarrow 19$, takes about 486 $\mu$s, or about 47 $\mu$s less than the transition from $v_X=81 \rightarrow 19$, resulting in a comparable total cascade time for the two initial levels. 
Similarly, for $v_X=55$, we obtain $\tau_\mathrm{cascade}$ of 36.06 s (via $v_X=11,9$), 21.68 s (via $v_X=2$), 31.93 s (via $v_X=23,22,4$), and and 30.09 s (via $v_X=23,20$).
The cascade times from the vibrational levels $v_X=18-20$, populated via $\mathrm{B}^1\Sigma^+$ state, will be up to about one second faster.

\section{Summary and Conclusions}
\label{sec:summary}

In this study, we have theoretically analyzed possible pathways for optical production of cold molecular ions in the ground electronic state from an atomic gas of ultracold Na interacting with a cold cloud of Ca$^{+}$ ions. 
To this end, we have calculated \textit{ab initio} electronic potential energy curves of (NaCa)$^{+}$ molecular ion for the ground and lowest eleven excited states of singlet symmetry, transition dipole moments between selected pairs of electronic states, and permanent dipole moments of the molecular ion in the two lowest-energy electronic states. The structure calculations were performed at a higher level of theory than reported in previous studies.

The computed electronic structure data, namely potentials and dipole moments, were used to analyze possible formation pathways involving single- and two-photon optical excitations. The analyzed formation schemes rely on photoassociation of the molecular ions in the $E^1\Sigma^+$ state as the first step, but differ in the approach to acheive population transfer to the ground electronic state.
For the experimental conditions feasible in current experiments, we predict that hundreds or thousands molecular ions per second could be photoassociated in the near-dissociation, loosely bound, vibrational levels in $E^1\Sigma^+$ state, that would decay spontaneously into highly excited vibrational levels of the $A^1\Sigma^+$ state, while a small fraction (less than 1\%) would decay directly into the ground electronic state. The PA properties are largely affected by the fact that the density of vibrational states near the dissociation threshold are governed by the long-range electronic potential, proportional to $1/R^4$ in the first order, resulting in very extended nature of such states when compared to the neutral dimers.

Consequently, the optimal approach would require a second excitation step to transfer the population to a spatially compact intermediate electronic state with better Franck-Condon overlap with the ground state. We find that both $B^1\Sigma^+$ and $C^1\Sigma^+$ states would be suitable candidates for the intermediate state, each with its own advantages and disadvantages. The $B^1\Sigma^+$ state offers better Franck-Condon overlap with lower vibrational levels in $X^1\Sigma^+$ state, requiring fewer spontaneous relaxation transitions to reach the lowest vibrational level, as well as minimal losses of population to spontaneous emission into higher excited electronic states. On the other hand, population transfer to the $C^1\Sigma^+$ state requires about 30 times smaller laser intensity $I_2$, which would allow a greater flexibility in the choice of vibrational transitions and experimental implementation. 
The proposed two-step process would yield molecular ions in excited vibrational levels of the ground state and would require approximately 40 seconds to decay into the lowest vibrational level. Such holding times should be possible to achieve in presently available experimental setups.

Alternatively, for a single-photon PA approach, we find that the molecular ions formed in low vibrational levels of the $E^1\Sigma^+$, such as $v_E=6$, can efficiently decay (up to 35\% of the population) directly into the $X^1\Sigma^+$ state, allowing for an efficient single-excitation production scheme. An implementation of such a setup would require higher PA laser intensities or higher densities of cold Ca$^+$ ions than assumed in our analysis. Alternatively, an enhanced PA scheme could be employed\cite{2008PhRvL.101e3201P}. 
Our production rates are several orders of magnitude higher than the radiative association rate, theoretically predicted to be $2.3\times10^{-16}$ cm$^{3}$s$^{-1}$ for Na and Ca$^+$ interacting on the A$^1\Sigma^+$ potential \cite{2003PhRvA..67d2705M}. Note that this rate would be further revised down if our transition dipole moment were used. The low value is consistent with measured radiative association rate reported in a hot sample \cite{2014ApPhB.114...75S}, although a direct comparison is not possible due to different entrance channels and the absence of spin-polarization in the experiment.
Radiative association rates of the order of $10^{-14}$ cm$^{3}$s$^{-1}$ to $10^{-16}$ cm$^{3}$s$^{-1}$ were predicted for other mixtures of cold atoms and ions (e.g. \cite{2015NJPh...17d5015D}), making the proposed PA mechanism worth considering in suitable atom-ion mixtures.
Finally, we note that triplet states were not included in the study since there is no experimental evidence for strong  singlet-triplet coupling of the considered electronic states \cite{2014ApPhB.114...75S}.

While the main goal of this study was to investigate the PA, as an all-optical approach to formation of ultracold (NaCa)$^+$ molecular ions, as well as to provide higher quality electronic structure data to aid future experiments, the results obtained suggest that a similar approach could be extended to other mixtures composed of an alkali metal atom and an alkaline earth ion with qualitatively similar electronic structure.
In fact, the electronic configuration of the excited states of (NaCa)$^+$ molecular ion does not make the short-range region readily accessible at low collision energies due to the presence of extended potential barriers in $^1\Sigma$ states, with an exception of $B^1\Sigma^+$ state, which, however, does not have a significant transition dipole moment with the $A^1\Sigma^+$ state. This is not the case in all heteronuclear mixtures where the conclusions of this study would be applicable, possibly resulting in larger transition rates in the short-range region and higher overall efficiency.

A possible extension of this study would require addition of the triplet $\Sigma$ and $\Pi$ electronic states and spin-orbit couplings in the analysis. The spin-orbit coupling could affect the transition rates in near-threshold ro-vibrational levels, resulting in changes of optimal formation pathways. Such analysis would ideally be done in conjunction with high-precision spectroscopy experiments. 
In addition, it would be interesting to analyze feasibility of the proposed approach in atom-ion mixtures at higher temperatures, where higher partial waves would contribute to the dynamics.

\begin{acknowledgments}
The authors wish to thank W.W. Smith for useful discussions. This work was partially supported by the MURI U.S. Army Research Office grant number W911NF-14-1-0378 (MG), and by the PIF program of the National Science Foundation grant number PHY-1415560 (RC).
\end{acknowledgments}

\bibliography{naca+_noarxiv}

\begin{thebibliography}{66}
\expandafter\ifx\csname natexlab\endcsname\relax\def\natexlab#1{#1}\fi
\expandafter\ifx\csname bibnamefont\endcsname\relax
  \def\bibnamefont#1{#1}\fi
\expandafter\ifx\csname bibfnamefont\endcsname\relax
  \def\bibfnamefont#1{#1}\fi
\expandafter\ifx\csname citenamefont\endcsname\relax
  \def\citenamefont#1{#1}\fi
\expandafter\ifx\csname url\endcsname\relax
  \def\url#1{\texttt{#1}}\fi
\expandafter\ifx\csname urlprefix\endcsname\relax\def\urlprefix{URL }\fi
\providecommand{\bibinfo}[2]{#2}
\providecommand{\eprint}[2][]{\url{#2}}

\bibitem[{\citenamefont{Krems et~al.}(2009)\citenamefont{Krems, Friedrich, and
  Stwalley}}]{krems2009cold}
\bibinfo{author}{\bibfnamefont{R.}~\bibnamefont{Krems}},
  \bibinfo{author}{\bibfnamefont{B.}~\bibnamefont{Friedrich}},
  \bibnamefont{and} \bibinfo{author}{\bibfnamefont{W.~C.}
  \bibnamefont{Stwalley}}, \emph{\bibinfo{title}{Cold molecules: theory,
  experiment, applications}} (\bibinfo{publisher}{CRC press},
  \bibinfo{year}{2009}).

\bibitem[{\citenamefont{{Carr} et~al.}(2009)\citenamefont{{Carr}, {DeMille},
  {Krems}, and {Ye}}}]{2009NJPh...11e5049C}
\bibinfo{author}{\bibfnamefont{L.~D.} \bibnamefont{{Carr}}},
  \bibinfo{author}{\bibfnamefont{D.}~\bibnamefont{{DeMille}}},
  \bibinfo{author}{\bibfnamefont{R.~V.} \bibnamefont{{Krems}}},
  \bibnamefont{and} \bibinfo{author}{\bibfnamefont{J.}~\bibnamefont{{Ye}}},
  \bibinfo{journal}{New J. Phys.} \textbf{\bibinfo{volume}{11}},
  \bibinfo{eid}{055049} (\bibinfo{year}{2009}).

\bibitem[{\citenamefont{{Blythe} et~al.}(2005)\citenamefont{{Blythe}, {Roth},
  {Fr{\"o}hlich}, {Wenz}, and {Schiller}}}]{2005PhRvL..95r3002B}
\bibinfo{author}{\bibfnamefont{P.}~\bibnamefont{{Blythe}}},
  \bibinfo{author}{\bibfnamefont{B.}~\bibnamefont{{Roth}}},
  \bibinfo{author}{\bibfnamefont{U.}~\bibnamefont{{Fr{\"o}hlich}}},
  \bibinfo{author}{\bibfnamefont{H.}~\bibnamefont{{Wenz}}}, \bibnamefont{and}
  \bibinfo{author}{\bibfnamefont{S.}~\bibnamefont{{Schiller}}},
  \bibinfo{journal}{\prl} \textbf{\bibinfo{volume}{95}}, \bibinfo{eid}{183002}
  (\bibinfo{year}{2005}).

\bibitem[{\citenamefont{{Hudson}}(2009)}]{2009PhRvA..79c2716H}
\bibinfo{author}{\bibfnamefont{E.~R.} \bibnamefont{{Hudson}}},
  \bibinfo{journal}{\pra} \textbf{\bibinfo{volume}{79}}, \bibinfo{eid}{032716}
  (\bibinfo{year}{2009}).

\bibitem[{\citenamefont{{Schneider}
  et~al.}(2010{\natexlab{a}})\citenamefont{{Schneider}, {Enderlein}, {Huber},
  and {Schaetz}}}]{2010NaPho...4..772S}
\bibinfo{author}{\bibfnamefont{C.}~\bibnamefont{{Schneider}}},
  \bibinfo{author}{\bibfnamefont{M.}~\bibnamefont{{Enderlein}}},
  \bibinfo{author}{\bibfnamefont{T.}~\bibnamefont{{Huber}}}, \bibnamefont{and}
  \bibinfo{author}{\bibfnamefont{T.}~\bibnamefont{{Schaetz}}},
  \bibinfo{journal}{Nat. Photon.} \textbf{\bibinfo{volume}{4}},
  \bibinfo{pages}{772} (\bibinfo{year}{2010}{\natexlab{a}}).

\bibitem[{\citenamefont{{Schneider}
  et~al.}(2010{\natexlab{b}})\citenamefont{{Schneider}, {Roth}, {Duncker},
  {Ernsting}, and {Schiller}}}]{2010NatPh...6..275S}
\bibinfo{author}{\bibfnamefont{T.}~\bibnamefont{{Schneider}}},
  \bibinfo{author}{\bibfnamefont{B.}~\bibnamefont{{Roth}}},
  \bibinfo{author}{\bibfnamefont{H.}~\bibnamefont{{Duncker}}},
  \bibinfo{author}{\bibfnamefont{I.}~\bibnamefont{{Ernsting}}},
  \bibnamefont{and}
  \bibinfo{author}{\bibfnamefont{S.}~\bibnamefont{{Schiller}}},
  \bibinfo{journal}{Nat. Phys.} \textbf{\bibinfo{volume}{6}},
  \bibinfo{pages}{275} (\bibinfo{year}{2010}{\natexlab{b}}).

\bibitem[{\citenamefont{{Huber} et~al.}(2014)\citenamefont{{Huber},
  {Lambrecht}, {Schmidt}, {Karpa}, and {Schaetz}}}]{2014NatCo...5E5587H}
\bibinfo{author}{\bibfnamefont{T.}~\bibnamefont{{Huber}}},
  \bibinfo{author}{\bibfnamefont{A.}~\bibnamefont{{Lambrecht}}},
  \bibinfo{author}{\bibfnamefont{J.}~\bibnamefont{{Schmidt}}},
  \bibinfo{author}{\bibfnamefont{L.}~\bibnamefont{{Karpa}}}, \bibnamefont{and}
  \bibinfo{author}{\bibfnamefont{T.}~\bibnamefont{{Schaetz}}},
  \bibinfo{journal}{Nat. Commun.} \textbf{\bibinfo{volume}{5}},
  \bibinfo{eid}{5587} (\bibinfo{year}{2014}).

\bibitem[{\citenamefont{{Krych} and {Idziaszek}}(2015)}]{2015PhRvA..91b3430K}
\bibinfo{author}{\bibfnamefont{M.}~\bibnamefont{{Krych}}} \bibnamefont{and}
  \bibinfo{author}{\bibfnamefont{Z.}~\bibnamefont{{Idziaszek}}},
  \bibinfo{journal}{\pra} \textbf{\bibinfo{volume}{91}}, \bibinfo{eid}{023430}
  (\bibinfo{year}{2015}).

\bibitem[{\citenamefont{{Lepers} et~al.}(2016)\citenamefont{{Lepers}, {Hong},
  {Wyart}, and {Dulieu}}}]{2016PhRvA..93a1401L}
\bibinfo{author}{\bibfnamefont{M.}~\bibnamefont{{Lepers}}},
  \bibinfo{author}{\bibfnamefont{Y.}~\bibnamefont{{Hong}}},
  \bibinfo{author}{\bibfnamefont{J.-F.} \bibnamefont{{Wyart}}},
  \bibnamefont{and} \bibinfo{author}{\bibfnamefont{O.}~\bibnamefont{{Dulieu}}},
  \bibinfo{journal}{\pra} \textbf{\bibinfo{volume}{93}}, \bibinfo{eid}{011401}
  (\bibinfo{year}{2016}).

\bibitem[{\citenamefont{{H{\"a}rter} and {Hecker
  Denschlag}}(2014)}]{2014ConPh..55...33H}
\bibinfo{author}{\bibfnamefont{A.}~\bibnamefont{{H{\"a}rter}}}
  \bibnamefont{and} \bibinfo{author}{\bibfnamefont{J.}~\bibnamefont{{Hecker
  Denschlag}}}, \bibinfo{journal}{Contemp. Phys.}
  \textbf{\bibinfo{volume}{55}}, \bibinfo{pages}{33} (\bibinfo{year}{2014}).

\bibitem[{\citenamefont{{Bissbort} et~al.}(2013)\citenamefont{{Bissbort},
  {Cocks}, {Negretti}, {Idziaszek}, {Calarco}, {Schmidt-Kaler}, {Hofstetter},
  and {Gerritsma}}}]{2013PhRvL.111h0501B}
\bibinfo{author}{\bibfnamefont{U.}~\bibnamefont{{Bissbort}}},
  \bibinfo{author}{\bibfnamefont{D.}~\bibnamefont{{Cocks}}},
  \bibinfo{author}{\bibfnamefont{A.}~\bibnamefont{{Negretti}}},
  \bibinfo{author}{\bibfnamefont{Z.}~\bibnamefont{{Idziaszek}}},
  \bibinfo{author}{\bibfnamefont{T.}~\bibnamefont{{Calarco}}},
  \bibinfo{author}{\bibfnamefont{F.}~\bibnamefont{{Schmidt-Kaler}}},
  \bibinfo{author}{\bibfnamefont{W.}~\bibnamefont{{Hofstetter}}},
  \bibnamefont{and}
  \bibinfo{author}{\bibfnamefont{R.}~\bibnamefont{{Gerritsma}}},
  \bibinfo{journal}{\prl} \textbf{\bibinfo{volume}{111}}, \bibinfo{eid}{080501}
  (\bibinfo{year}{2013}).

\bibitem[{\citenamefont{{Gerritsma} et~al.}(2012)\citenamefont{{Gerritsma},
  {Negretti}, {Doerk}, {Idziaszek}, {Calarco}, and
  {Schmidt-Kaler}}}]{2012PhRvL.109h0402G}
\bibinfo{author}{\bibfnamefont{R.}~\bibnamefont{{Gerritsma}}},
  \bibinfo{author}{\bibfnamefont{A.}~\bibnamefont{{Negretti}}},
  \bibinfo{author}{\bibfnamefont{H.}~\bibnamefont{{Doerk}}},
  \bibinfo{author}{\bibfnamefont{Z.}~\bibnamefont{{Idziaszek}}},
  \bibinfo{author}{\bibfnamefont{T.}~\bibnamefont{{Calarco}}},
  \bibnamefont{and}
  \bibinfo{author}{\bibfnamefont{F.}~\bibnamefont{{Schmidt-Kaler}}},
  \bibinfo{journal}{\prl} \textbf{\bibinfo{volume}{109}}, \bibinfo{eid}{080402}
  (\bibinfo{year}{2012}).

\bibitem[{\citenamefont{{C{\^o}t{\'e}}}(2000)}]{2000PhRvL..85.5316C}
\bibinfo{author}{\bibfnamefont{R.}~\bibnamefont{{C{\^o}t{\'e}}}},
  \bibinfo{journal}{\prl} \textbf{\bibinfo{volume}{85}}, \bibinfo{pages}{5316}
  (\bibinfo{year}{2000}).

\bibitem[{\citenamefont{{C{\^o}t{\'e}}
  et~al.}(2002)\citenamefont{{C{\^o}t{\'e}}, {Kharchenko}, and
  {Lukin}}}]{2002PhRvL..89i3001C}
\bibinfo{author}{\bibfnamefont{R.}~\bibnamefont{{C{\^o}t{\'e}}}},
  \bibinfo{author}{\bibfnamefont{V.}~\bibnamefont{{Kharchenko}}},
  \bibnamefont{and} \bibinfo{author}{\bibfnamefont{M.~D.}
  \bibnamefont{{Lukin}}}, \bibinfo{journal}{\prl}
  \textbf{\bibinfo{volume}{89}}, \bibinfo{eid}{093001} (\bibinfo{year}{2002}).

\bibitem[{\citenamefont{{Cucchietti} and
  {Timmermans}}(2006)}]{2006PhRvL..96u0401C}
\bibinfo{author}{\bibfnamefont{F.~M.} \bibnamefont{{Cucchietti}}}
  \bibnamefont{and}
  \bibinfo{author}{\bibfnamefont{E.}~\bibnamefont{{Timmermans}}},
  \bibinfo{journal}{\prl} \textbf{\bibinfo{volume}{96}}, \bibinfo{eid}{210401}
  (\bibinfo{year}{2006}).

\bibitem[{\citenamefont{{Balewski} et~al.}(2013)\citenamefont{{Balewski},
  {Krupp}, {Gaj}, {Peter}, {B{\"u}chler}, {L{\"o}w}, {Hofferberth}, and
  {Pfau}}}]{2013Natur.502..664B}
\bibinfo{author}{\bibfnamefont{J.~B.} \bibnamefont{{Balewski}}},
  \bibinfo{author}{\bibfnamefont{A.~T.} \bibnamefont{{Krupp}}},
  \bibinfo{author}{\bibfnamefont{A.}~\bibnamefont{{Gaj}}},
  \bibinfo{author}{\bibfnamefont{D.}~\bibnamefont{{Peter}}},
  \bibinfo{author}{\bibfnamefont{H.~P.} \bibnamefont{{B{\"u}chler}}},
  \bibinfo{author}{\bibfnamefont{R.}~\bibnamefont{{L{\"o}w}}},
  \bibinfo{author}{\bibfnamefont{S.}~\bibnamefont{{Hofferberth}}},
  \bibnamefont{and} \bibinfo{author}{\bibfnamefont{T.}~\bibnamefont{{Pfau}}},
  \bibinfo{journal}{\nat} \textbf{\bibinfo{volume}{502}}, \bibinfo{pages}{664}
  (\bibinfo{year}{2013}).

\bibitem[{\citenamefont{{Wang} et~al.}(2015)\citenamefont{{Wang}, {Gacesa}, and
  {C{\^o}t{\'e}}}}]{2015PhRvL.114x3003W}
\bibinfo{author}{\bibfnamefont{J.}~\bibnamefont{{Wang}}},
  \bibinfo{author}{\bibfnamefont{M.}~\bibnamefont{{Gacesa}}}, \bibnamefont{and}
  \bibinfo{author}{\bibfnamefont{R.}~\bibnamefont{{C{\^o}t{\'e}}}},
  \bibinfo{journal}{\prl} \textbf{\bibinfo{volume}{114}}, \bibinfo{eid}{243003}
  (\bibinfo{year}{2015}).

\bibitem[{\citenamefont{{Doerk} et~al.}(2010)\citenamefont{{Doerk},
  {Idziaszek}, and {Calarco}}}]{2010PhRvA..81a2708D}
\bibinfo{author}{\bibfnamefont{H.}~\bibnamefont{{Doerk}}},
  \bibinfo{author}{\bibfnamefont{Z.}~\bibnamefont{{Idziaszek}}},
  \bibnamefont{and}
  \bibinfo{author}{\bibfnamefont{T.}~\bibnamefont{{Calarco}}},
  \bibinfo{journal}{\pra} \textbf{\bibinfo{volume}{81}}, \bibinfo{eid}{012708}
  (\bibinfo{year}{2010}).

\bibitem[{\citenamefont{{Kuznetsova} et~al.}(2010)\citenamefont{{Kuznetsova},
  {Gacesa}, {Yelin}, and {C{\^o}t{\'e}}}}]{2010PhRvA..81c0301K}
\bibinfo{author}{\bibfnamefont{E.}~\bibnamefont{{Kuznetsova}}},
  \bibinfo{author}{\bibfnamefont{M.}~\bibnamefont{{Gacesa}}},
  \bibinfo{author}{\bibfnamefont{S.~F.} \bibnamefont{{Yelin}}},
  \bibnamefont{and}
  \bibinfo{author}{\bibfnamefont{R.}~\bibnamefont{{C{\^o}t{\'e}}}},
  \bibinfo{journal}{\pra} \textbf{\bibinfo{volume}{81}}, \bibinfo{eid}{030301}
  (\bibinfo{year}{2010}).

\bibitem[{\citenamefont{{C{\^o}t{\'e}} and
  {Dalgarno}}(2000)}]{2000PhRvA..62a2709C}
\bibinfo{author}{\bibfnamefont{R.}~\bibnamefont{{C{\^o}t{\'e}}}}
  \bibnamefont{and}
  \bibinfo{author}{\bibfnamefont{A.}~\bibnamefont{{Dalgarno}}},
  \bibinfo{journal}{\pra} \textbf{\bibinfo{volume}{62}}, \bibinfo{eid}{012709}
  (\bibinfo{year}{2000}).

\bibitem[{\citenamefont{{Gao}}(2011)}]{2011PhRvA..83f2712G}
\bibinfo{author}{\bibfnamefont{B.}~\bibnamefont{{Gao}}},
  \bibinfo{journal}{\pra} \textbf{\bibinfo{volume}{83}}, \bibinfo{eid}{062712}
  (\bibinfo{year}{2011}).

\bibitem[{\citenamefont{{Gao}}(2013)}]{2013PhRvA..88b2701G}
\bibinfo{author}{\bibfnamefont{B.}~\bibnamefont{{Gao}}},
  \bibinfo{journal}{\pra} \textbf{\bibinfo{volume}{88}}, \bibinfo{eid}{022701}
  (\bibinfo{year}{2013}).

\bibitem[{\citenamefont{{P{\'e}rez-R{\'{\i}}os} and
  {Greene}}(2015)}]{2015JChPh.143d1105P}
\bibinfo{author}{\bibfnamefont{J.}~\bibnamefont{{P{\'e}rez-R{\'{\i}}os}}}
  \bibnamefont{and} \bibinfo{author}{\bibfnamefont{C.~H.}
  \bibnamefont{{Greene}}}, \bibinfo{journal}{\jcp}
  \textbf{\bibinfo{volume}{143}}, \bibinfo{eid}{041105} (\bibinfo{year}{2015}).

\bibitem[{\citenamefont{{Cetina} et~al.}(2012)\citenamefont{{Cetina}, {Grier},
  and {Vuleti{\'c}}}}]{2012PhRvL.109y3201C}
\bibinfo{author}{\bibfnamefont{M.}~\bibnamefont{{Cetina}}},
  \bibinfo{author}{\bibfnamefont{A.~T.} \bibnamefont{{Grier}}},
  \bibnamefont{and}
  \bibinfo{author}{\bibfnamefont{V.}~\bibnamefont{{Vuleti{\'c}}}},
  \bibinfo{journal}{\prl} \textbf{\bibinfo{volume}{109}}, \bibinfo{eid}{253201}
  (\bibinfo{year}{2012}).

\bibitem[{\citenamefont{{Ravi} et~al.}(2012)\citenamefont{{Ravi}, {Lee},
  {Sharma}, {Werth}, and {Rangwala}}}]{2012NatCo...3E1126R}
\bibinfo{author}{\bibfnamefont{K.}~\bibnamefont{{Ravi}}},
  \bibinfo{author}{\bibfnamefont{S.}~\bibnamefont{{Lee}}},
  \bibinfo{author}{\bibfnamefont{A.}~\bibnamefont{{Sharma}}},
  \bibinfo{author}{\bibfnamefont{G.}~\bibnamefont{{Werth}}}, \bibnamefont{and}
  \bibinfo{author}{\bibfnamefont{S.~A.} \bibnamefont{{Rangwala}}},
  \bibinfo{journal}{Nat. Commun.} \textbf{\bibinfo{volume}{3}},
  \bibinfo{eid}{1126} (\bibinfo{year}{2012}).

\bibitem[{\citenamefont{{Hall} et~al.}(2011)\citenamefont{{Hall}, {Aymar},
  {Bouloufa-Maafa}, {Dulieu}, and {Willitsch}}}]{2011PhRvL.107x3202H}
\bibinfo{author}{\bibfnamefont{F.~H.~J.} \bibnamefont{{Hall}}},
  \bibinfo{author}{\bibfnamefont{M.}~\bibnamefont{{Aymar}}},
  \bibinfo{author}{\bibfnamefont{N.}~\bibnamefont{{Bouloufa-Maafa}}},
  \bibinfo{author}{\bibfnamefont{O.}~\bibnamefont{{Dulieu}}}, \bibnamefont{and}
  \bibinfo{author}{\bibfnamefont{S.}~\bibnamefont{{Willitsch}}},
  \bibinfo{journal}{\prl} \textbf{\bibinfo{volume}{107}}, \bibinfo{eid}{243202}
  (\bibinfo{year}{2011}).

\bibitem[{\citenamefont{{Hall} et~al.}(2013{\natexlab{a}})\citenamefont{{Hall},
  {Eberle}, {Hegi}, {Raoult}, {Aymar}, {Dulieu}, and
  {Willitsch}}}]{2013MolPh.111.2020H}
\bibinfo{author}{\bibfnamefont{F.~H.~J.} \bibnamefont{{Hall}}},
  \bibinfo{author}{\bibfnamefont{P.}~\bibnamefont{{Eberle}}},
  \bibinfo{author}{\bibfnamefont{G.}~\bibnamefont{{Hegi}}},
  \bibinfo{author}{\bibfnamefont{M.}~\bibnamefont{{Raoult}}},
  \bibinfo{author}{\bibfnamefont{M.}~\bibnamefont{{Aymar}}},
  \bibinfo{author}{\bibfnamefont{O.}~\bibnamefont{{Dulieu}}}, \bibnamefont{and}
  \bibinfo{author}{\bibfnamefont{S.}~\bibnamefont{{Willitsch}}},
  \bibinfo{journal}{Mol. Phys.} \textbf{\bibinfo{volume}{111}},
  \bibinfo{pages}{2020} (\bibinfo{year}{2013}{\natexlab{a}}).

\bibitem[{\citenamefont{{Hall} et~al.}(2013{\natexlab{b}})\citenamefont{{Hall},
  {Aymar}, {Raoult}, {Dulieu}, and {Willitsch}}}]{2013MolPh.111.1683H}
\bibinfo{author}{\bibfnamefont{F.~H.~J.} \bibnamefont{{Hall}}},
  \bibinfo{author}{\bibfnamefont{M.}~\bibnamefont{{Aymar}}},
  \bibinfo{author}{\bibfnamefont{M.}~\bibnamefont{{Raoult}}},
  \bibinfo{author}{\bibfnamefont{O.}~\bibnamefont{{Dulieu}}}, \bibnamefont{and}
  \bibinfo{author}{\bibfnamefont{S.}~\bibnamefont{{Willitsch}}},
  \bibinfo{journal}{Mol. Phys.} \textbf{\bibinfo{volume}{111}},
  \bibinfo{pages}{1683} (\bibinfo{year}{2013}{\natexlab{b}}).

\bibitem[{\citenamefont{{Bodo} et~al.}(2008)\citenamefont{{Bodo}, {Zhang}, and
  {Dalgarno}}}]{2008NJPh...10c3024B}
\bibinfo{author}{\bibfnamefont{E.}~\bibnamefont{{Bodo}}},
  \bibinfo{author}{\bibfnamefont{P.}~\bibnamefont{{Zhang}}}, \bibnamefont{and}
  \bibinfo{author}{\bibfnamefont{A.}~\bibnamefont{{Dalgarno}}},
  \bibinfo{journal}{New J. Phys.} \textbf{\bibinfo{volume}{10}},
  \bibinfo{eid}{033024} (\bibinfo{year}{2008}).

\bibitem[{\citenamefont{{Yan} et~al.}(2013)\citenamefont{{Yan}, {Liu}, {Wu},
  {Qu}, {Wang}, and {Buenker}}}]{2013PhRvA..88a2709Y}
\bibinfo{author}{\bibfnamefont{L.~L.} \bibnamefont{{Yan}}},
  \bibinfo{author}{\bibfnamefont{L.}~\bibnamefont{{Liu}}},
  \bibinfo{author}{\bibfnamefont{Y.}~\bibnamefont{{Wu}}},
  \bibinfo{author}{\bibfnamefont{Y.~Z.} \bibnamefont{{Qu}}},
  \bibinfo{author}{\bibfnamefont{J.~G.} \bibnamefont{{Wang}}},
  \bibnamefont{and} \bibinfo{author}{\bibfnamefont{R.~J.}
  \bibnamefont{{Buenker}}}, \bibinfo{journal}{\pra}
  \textbf{\bibinfo{volume}{88}}, \bibinfo{eid}{012709} (\bibinfo{year}{2013}).

\bibitem[{\citenamefont{{Yan} et~al.}(2014)\citenamefont{{Yan}, {Li}, {Wu},
  {Wang}, and {Qu}}}]{2014PhRvA..90c2714Y}
\bibinfo{author}{\bibfnamefont{L.~L.} \bibnamefont{{Yan}}},
  \bibinfo{author}{\bibfnamefont{X.~Y.} \bibnamefont{{Li}}},
  \bibinfo{author}{\bibfnamefont{Y.}~\bibnamefont{{Wu}}},
  \bibinfo{author}{\bibfnamefont{J.~G.} \bibnamefont{{Wang}}},
  \bibnamefont{and} \bibinfo{author}{\bibfnamefont{Y.~Z.} \bibnamefont{{Qu}}},
  \bibinfo{journal}{\pra} \textbf{\bibinfo{volume}{90}}, \bibinfo{eid}{032714}
  (\bibinfo{year}{2014}).

\bibitem[{\citenamefont{{Tomza} et~al.}(2015)\citenamefont{{Tomza}, {Koch}, and
  {Moszynski}}}]{2015PhRvA..91d2706T}
\bibinfo{author}{\bibfnamefont{M.}~\bibnamefont{{Tomza}}},
  \bibinfo{author}{\bibfnamefont{C.~P.} \bibnamefont{{Koch}}},
  \bibnamefont{and}
  \bibinfo{author}{\bibfnamefont{R.}~\bibnamefont{{Moszynski}}},
  \bibinfo{journal}{\pra} \textbf{\bibinfo{volume}{91}}, \bibinfo{eid}{042706}
  (\bibinfo{year}{2015}).

\bibitem[{\citenamefont{{Sayfutyarova}
  et~al.}(2013)\citenamefont{{Sayfutyarova}, {Buchachenko}, {Yakovleva}, and
  {Belyaev}}}]{2013PhRvA..87e2717S}
\bibinfo{author}{\bibfnamefont{E.~R.} \bibnamefont{{Sayfutyarova}}},
  \bibinfo{author}{\bibfnamefont{A.~A.} \bibnamefont{{Buchachenko}}},
  \bibinfo{author}{\bibfnamefont{S.~A.} \bibnamefont{{Yakovleva}}},
  \bibnamefont{and} \bibinfo{author}{\bibfnamefont{A.~K.}
  \bibnamefont{{Belyaev}}}, \bibinfo{journal}{\pra}
  \textbf{\bibinfo{volume}{87}}, \bibinfo{eid}{052717} (\bibinfo{year}{2013}).

\bibitem[{\citenamefont{{da Silva} et~al.}(2015)\citenamefont{{da Silva},
  {Raoult}, {Aymar}, and {Dulieu}}}]{2015NJPh...17d5015D}
\bibinfo{author}{\bibfnamefont{H.}~\bibnamefont{{da Silva}},
  \bibfnamefont{Jr.}},
  \bibinfo{author}{\bibfnamefont{M.}~\bibnamefont{{Raoult}}},
  \bibinfo{author}{\bibfnamefont{M.}~\bibnamefont{{Aymar}}}, \bibnamefont{and}
  \bibinfo{author}{\bibfnamefont{O.}~\bibnamefont{{Dulieu}}},
  \bibinfo{journal}{New J. Phys.} \textbf{\bibinfo{volume}{17}},
  \bibinfo{eid}{045015} (\bibinfo{year}{2015}).

\bibitem[{\citenamefont{{Rakshit} and {Deb}}(2011)}]{2011PhRvA..83b2703R}
\bibinfo{author}{\bibfnamefont{A.}~\bibnamefont{{Rakshit}}} \bibnamefont{and}
  \bibinfo{author}{\bibfnamefont{B.}~\bibnamefont{{Deb}}},
  \bibinfo{journal}{\pra} \textbf{\bibinfo{volume}{83}}, \bibinfo{eid}{022703}
  (\bibinfo{year}{2011}).

\bibitem[{\citenamefont{{Rellergert} et~al.}(2011)\citenamefont{{Rellergert},
  {Sullivan}, {Kotochigova}, {Petrov}, {Chen}, {Schowalter}, and
  {Hudson}}}]{2011PhRvL.107x3201R}
\bibinfo{author}{\bibfnamefont{W.~G.} \bibnamefont{{Rellergert}}},
  \bibinfo{author}{\bibfnamefont{S.~T.} \bibnamefont{{Sullivan}}},
  \bibinfo{author}{\bibfnamefont{S.}~\bibnamefont{{Kotochigova}}},
  \bibinfo{author}{\bibfnamefont{A.}~\bibnamefont{{Petrov}}},
  \bibinfo{author}{\bibfnamefont{K.}~\bibnamefont{{Chen}}},
  \bibinfo{author}{\bibfnamefont{S.~J.} \bibnamefont{{Schowalter}}},
  \bibnamefont{and} \bibinfo{author}{\bibfnamefont{E.~R.}
  \bibnamefont{{Hudson}}}, \bibinfo{journal}{\prl}
  \textbf{\bibinfo{volume}{107}}, \bibinfo{eid}{243201} (\bibinfo{year}{2011}).

\bibitem[{\citenamefont{{Zygelman} et~al.}(2014)\citenamefont{{Zygelman},
  {Lucic}, and {Hudson}}}]{2014JPhB...47a5301Z}
\bibinfo{author}{\bibfnamefont{B.}~\bibnamefont{{Zygelman}}},
  \bibinfo{author}{\bibfnamefont{Z.}~\bibnamefont{{Lucic}}}, \bibnamefont{and}
  \bibinfo{author}{\bibfnamefont{E.~R.} \bibnamefont{{Hudson}}},
  \bibinfo{journal}{J. Phys. B} \textbf{\bibinfo{volume}{47}},
  \bibinfo{eid}{015301} (\bibinfo{year}{2014}).

\bibitem[{\citenamefont{{Goodman} et~al.}(2015)\citenamefont{{Goodman},
  {Wells}, {Kwolek}, {Bl{\"u}mel}, {Narducci}, and
  {Smith}}}]{2015PhRvA..91a2709G}
\bibinfo{author}{\bibfnamefont{D.~S.} \bibnamefont{{Goodman}}},
  \bibinfo{author}{\bibfnamefont{J.~E.} \bibnamefont{{Wells}}},
  \bibinfo{author}{\bibfnamefont{J.~M.} \bibnamefont{{Kwolek}}},
  \bibinfo{author}{\bibfnamefont{R.}~\bibnamefont{{Bl{\"u}mel}}},
  \bibinfo{author}{\bibfnamefont{F.~A.} \bibnamefont{{Narducci}}},
  \bibnamefont{and} \bibinfo{author}{\bibfnamefont{W.~W.}
  \bibnamefont{{Smith}}}, \bibinfo{journal}{\pra}
  \textbf{\bibinfo{volume}{91}}, \bibinfo{eid}{012709} (\bibinfo{year}{2015}).

\bibitem[{\citenamefont{{Makarov} et~al.}(2003)\citenamefont{{Makarov},
  {C{\^o}t{\'e}}, {Michels}, and {Smith}}}]{2003PhRvA..67d2705M}
\bibinfo{author}{\bibfnamefont{O.~P.} \bibnamefont{{Makarov}}},
  \bibinfo{author}{\bibfnamefont{R.}~\bibnamefont{{C{\^o}t{\'e}}}},
  \bibinfo{author}{\bibfnamefont{H.}~\bibnamefont{{Michels}}},
  \bibnamefont{and} \bibinfo{author}{\bibfnamefont{W.~W.}
  \bibnamefont{{Smith}}}, \bibinfo{journal}{\pra}
  \textbf{\bibinfo{volume}{67}}, \bibinfo{eid}{042705} (\bibinfo{year}{2003}).

\bibitem[{\citenamefont{{Smith} et~al.}(2014)\citenamefont{{Smith}, {Goodman},
  {Sivarajah}, {Wells}, {Banerjee}, {C{\^o}t{\'e}}, {Michels}, {Mongtomery},
  and {Narducci}}}]{2014ApPhB.114...75S}
\bibinfo{author}{\bibfnamefont{W.~W.} \bibnamefont{{Smith}}},
  \bibinfo{author}{\bibfnamefont{D.~S.} \bibnamefont{{Goodman}}},
  \bibinfo{author}{\bibfnamefont{I.}~\bibnamefont{{Sivarajah}}},
  \bibinfo{author}{\bibfnamefont{J.~E.} \bibnamefont{{Wells}}},
  \bibinfo{author}{\bibfnamefont{S.}~\bibnamefont{{Banerjee}}},
  \bibinfo{author}{\bibfnamefont{R.}~\bibnamefont{{C{\^o}t{\'e}}}},
  \bibinfo{author}{\bibfnamefont{H.~H.} \bibnamefont{{Michels}}},
  \bibinfo{author}{\bibfnamefont{J.~A.} \bibnamefont{{Mongtomery}}},
  \bibnamefont{and} \bibinfo{author}{\bibfnamefont{F.~A.}
  \bibnamefont{{Narducci}}}, \bibinfo{journal}{Appl. Phys. B}
  \textbf{\bibinfo{volume}{114}}, \bibinfo{pages}{75} (\bibinfo{year}{2014}).

\bibitem[{\citenamefont{{Zhang} et~al.}(2011)\citenamefont{{Zhang}, {Dalgarno},
  {C{\^o}t{\'e}}, and {Bodo}}}]{2011PCCP...1319026Z}
\bibinfo{author}{\bibfnamefont{P.}~\bibnamefont{{Zhang}}},
  \bibinfo{author}{\bibfnamefont{A.}~\bibnamefont{{Dalgarno}}},
  \bibinfo{author}{\bibfnamefont{R.}~\bibnamefont{{C{\^o}t{\'e}}}},
  \bibnamefont{and} \bibinfo{author}{\bibfnamefont{E.}~\bibnamefont{{Bodo}}},
  \bibinfo{journal}{Phys. Chem. Chem. Phys.} \textbf{\bibinfo{volume}{13}},
  \bibinfo{pages}{19026} (\bibinfo{year}{2011}).

\bibitem[{\citenamefont{{Grier} et~al.}(2009)\citenamefont{{Grier}, {Cetina},
  {Oru{\v c}evi{\'c}}, and {Vuleti{\'c}}}}]{2009PhRvL.102v3201G}
\bibinfo{author}{\bibfnamefont{A.~T.} \bibnamefont{{Grier}}},
  \bibinfo{author}{\bibfnamefont{M.}~\bibnamefont{{Cetina}}},
  \bibinfo{author}{\bibfnamefont{F.}~\bibnamefont{{Oru{\v c}evi{\'c}}}},
  \bibnamefont{and}
  \bibinfo{author}{\bibfnamefont{V.}~\bibnamefont{{Vuleti{\'c}}}},
  \bibinfo{journal}{\prl} \textbf{\bibinfo{volume}{102}}, \bibinfo{eid}{223201}
  (\bibinfo{year}{2009}).

\bibitem[{\citenamefont{{McLaughlin} et~al.}(2014)\citenamefont{{McLaughlin},
  {Lamb}, {Lane}, and {McCann}}}]{2014JPhB...47n5201M}
\bibinfo{author}{\bibfnamefont{B.~M.} \bibnamefont{{McLaughlin}}},
  \bibinfo{author}{\bibfnamefont{H.~D.~L.} \bibnamefont{{Lamb}}},
  \bibinfo{author}{\bibfnamefont{I.~C.} \bibnamefont{{Lane}}},
  \bibnamefont{and} \bibinfo{author}{\bibfnamefont{J.~F.}
  \bibnamefont{{McCann}}}, \bibinfo{journal}{J. Phys. B}
  \textbf{\bibinfo{volume}{47}}, \bibinfo{eid}{145201} (\bibinfo{year}{2014}).

\bibitem[{\citenamefont{{Li} et~al.}(2015)\citenamefont{{Li}, {Qu}, {Wu},
  {Liu}, {Wang}, {Liebermann}, and {Buenker}}}]{2015PhRvA..91e2702L}
\bibinfo{author}{\bibfnamefont{T.~C.} \bibnamefont{{Li}}},
  \bibinfo{author}{\bibfnamefont{Y.~Z.} \bibnamefont{{Qu}}},
  \bibinfo{author}{\bibfnamefont{Y.}~\bibnamefont{{Wu}}},
  \bibinfo{author}{\bibfnamefont{L.}~\bibnamefont{{Liu}}},
  \bibinfo{author}{\bibfnamefont{J.~G.} \bibnamefont{{Wang}}},
  \bibinfo{author}{\bibfnamefont{H.-P.} \bibnamefont{{Liebermann}}},
  \bibnamefont{and} \bibinfo{author}{\bibfnamefont{R.~J.}
  \bibnamefont{{Buenker}}}, \bibinfo{journal}{\pra}
  \textbf{\bibinfo{volume}{91}}, \bibinfo{eid}{052702} (\bibinfo{year}{2015}).

\bibitem[{\citenamefont{{Rakshit} et~al.}(2015)\citenamefont{{Rakshit},
  {Ghanmi}, {Berriche}, and {Deb}}}]{2015arXiv150403114R}
\bibinfo{author}{\bibfnamefont{A.}~\bibnamefont{{Rakshit}}},
  \bibinfo{author}{\bibfnamefont{C.}~\bibnamefont{{Ghanmi}}},
  \bibinfo{author}{\bibfnamefont{H.}~\bibnamefont{{Berriche}}},
  \bibnamefont{and} \bibinfo{author}{\bibfnamefont{B.}~\bibnamefont{{Deb}}},
  \bibinfo{journal}{ArXiv e-prints}  (\bibinfo{year}{2015}),
  \eprint{1504.03114}.

\bibitem[{\citenamefont{{Thorsheim} et~al.}(1987)\citenamefont{{Thorsheim},
  {Weiner}, and {Julienne}}}]{1987PhRvL..58.2420T}
\bibinfo{author}{\bibfnamefont{H.~R.} \bibnamefont{{Thorsheim}}},
  \bibinfo{author}{\bibfnamefont{J.}~\bibnamefont{{Weiner}}}, \bibnamefont{and}
  \bibinfo{author}{\bibfnamefont{P.~S.} \bibnamefont{{Julienne}}},
  \bibinfo{journal}{\prl} \textbf{\bibinfo{volume}{58}}, \bibinfo{pages}{2420}
  (\bibinfo{year}{1987}).

\bibitem[{\citenamefont{Jones et~al.}(2006)\citenamefont{Jones, Tiesinga, Lett,
  and Julienne}}]{RevModPhys.78.483}
\bibinfo{author}{\bibfnamefont{K.~M.} \bibnamefont{Jones}},
  \bibinfo{author}{\bibfnamefont{E.}~\bibnamefont{Tiesinga}},
  \bibinfo{author}{\bibfnamefont{P.~D.} \bibnamefont{Lett}}, \bibnamefont{and}
  \bibinfo{author}{\bibfnamefont{P.~S.} \bibnamefont{Julienne}},
  \bibinfo{journal}{Rev. Mod. Phys.} \textbf{\bibinfo{volume}{78}},
  \bibinfo{pages}{483} (\bibinfo{year}{2006}).

\bibitem[{\citenamefont{Stwalley and
  Wang}(1999)}]{stwalley1999photoassociation}
\bibinfo{author}{\bibfnamefont{W.~C.} \bibnamefont{Stwalley}} \bibnamefont{and}
  \bibinfo{author}{\bibfnamefont{H.}~\bibnamefont{Wang}}, \bibinfo{journal}{J.
  Mol. Spectr.} \textbf{\bibinfo{volume}{195}}, \bibinfo{pages}{194}
  (\bibinfo{year}{1999}).

\bibitem[{\citenamefont{Stanton and Bartlett}(1993)}]{Stanton_Bartlett_1993}
\bibinfo{author}{\bibfnamefont{J.~F.} \bibnamefont{Stanton}} \bibnamefont{and}
  \bibinfo{author}{\bibfnamefont{R.~J.} \bibnamefont{Bartlett}},
  \bibinfo{journal}{J. Chem. Phys.} \textbf{\bibinfo{volume}{98}}
  (\bibinfo{year}{1993}).

\bibitem[{\citenamefont{{Korona} and {Werner}}(2003)}]{2003JChPh.118.3006K}
\bibinfo{author}{\bibfnamefont{T.}~\bibnamefont{{Korona}}} \bibnamefont{and}
  \bibinfo{author}{\bibfnamefont{H.-J.} \bibnamefont{{Werner}}},
  \bibinfo{journal}{\jcp} \textbf{\bibinfo{volume}{118}}, \bibinfo{pages}{3006}
  (\bibinfo{year}{2003}).

\bibitem[{\citenamefont{Werner et~al.}(2012)\citenamefont{Werner, Knowles,
  Knizia, Manby, {Sch\"{u}tz}, Celani, Korona, Lindh, Mitrushenkov, Rauhut
  et~al.}}]{MOLPRO}
\bibinfo{author}{\bibfnamefont{H.-J.} \bibnamefont{Werner}},
  \bibinfo{author}{\bibfnamefont{P.~J.} \bibnamefont{Knowles}},
  \bibinfo{author}{\bibfnamefont{G.}~\bibnamefont{Knizia}},
  \bibinfo{author}{\bibfnamefont{F.~R.} \bibnamefont{Manby}},
  \bibinfo{author}{\bibfnamefont{M.}~\bibnamefont{{Sch\"{u}tz}}},
  \bibinfo{author}{\bibfnamefont{P.}~\bibnamefont{Celani}},
  \bibinfo{author}{\bibfnamefont{T.}~\bibnamefont{Korona}},
  \bibinfo{author}{\bibfnamefont{R.}~\bibnamefont{Lindh}},
  \bibinfo{author}{\bibfnamefont{A.}~\bibnamefont{Mitrushenkov}},
  \bibinfo{author}{\bibfnamefont{G.}~\bibnamefont{Rauhut}},
  \bibnamefont{et~al.}, \emph{\bibinfo{title}{{MOLPRO}, v2012.1, a package of
  ab initio programs}} (\bibinfo{year}{2012}), \bibinfo{note}{see
  http://www.molpro.net}.

\bibitem[{\citenamefont{Fuentealba et~al.}(1982)\citenamefont{Fuentealba,
  Preuss, Stoll, and v.~Szentpaly}}]{Fuentealba1982}
\bibinfo{author}{\bibfnamefont{P.}~\bibnamefont{Fuentealba}},
  \bibinfo{author}{\bibfnamefont{H.}~\bibnamefont{Preuss}},
  \bibinfo{author}{\bibfnamefont{H.}~\bibnamefont{Stoll}}, \bibnamefont{and}
  \bibinfo{author}{\bibfnamefont{L.}~\bibnamefont{v.~Szentpaly}},
  \bibinfo{journal}{Chem. Phys. Lett.} \textbf{\bibinfo{volume}{89}},
  \bibinfo{pages}{418} (\bibinfo{year}{1982}).

\bibitem[{\citenamefont{Fuentealba et~al.}(1985)\citenamefont{Fuentealba, von
  Szentpaly, Preuss, and Stoll}}]{Fuentealba1985}
\bibinfo{author}{\bibfnamefont{P.}~\bibnamefont{Fuentealba}},
  \bibinfo{author}{\bibfnamefont{L.}~\bibnamefont{von Szentpaly}},
  \bibinfo{author}{\bibfnamefont{H.}~\bibnamefont{Preuss}}, \bibnamefont{and}
  \bibinfo{author}{\bibfnamefont{H.}~\bibnamefont{Stoll}}, \bibinfo{journal}{J.
  Phys. B} \textbf{\bibinfo{volume}{18}}, \bibinfo{pages}{1287}
  (\bibinfo{year}{1985}).

\bibitem[{\citenamefont{{Aymar} and {Dulieu}}(2005)}]{2005JChPh.122t4302A}
\bibinfo{author}{\bibfnamefont{M.}~\bibnamefont{{Aymar}}} \bibnamefont{and}
  \bibinfo{author}{\bibfnamefont{O.}~\bibnamefont{{Dulieu}}},
  \bibinfo{journal}{\jcp} \textbf{\bibinfo{volume}{122}},
  \bibinfo{pages}{204302} (\bibinfo{year}{2005}).

\bibitem[{\citenamefont{Czuchaj et~al.}(2003)\citenamefont{Czuchaj,
  Kro{\'s}nicki, and Stoll}}]{czuchaj2003valence}
\bibinfo{author}{\bibfnamefont{E.}~\bibnamefont{Czuchaj}},
  \bibinfo{author}{\bibfnamefont{M.}~\bibnamefont{Kro{\'s}nicki}},
  \bibnamefont{and} \bibinfo{author}{\bibfnamefont{H.}~\bibnamefont{Stoll}},
  \bibinfo{journal}{Theor. Chem. Acc.} \textbf{\bibinfo{volume}{110}},
  \bibinfo{pages}{28} (\bibinfo{year}{2003}).

\bibitem[{\citenamefont{{Banerjee} et~al.}(2012)\citenamefont{{Banerjee},
  {Montgomery}, {Byrd}, {Michels}, and {C{\^o}t{\'e}}}}]{2012CPL...542..138B}
\bibinfo{author}{\bibfnamefont{S.}~\bibnamefont{{Banerjee}}},
  \bibinfo{author}{\bibfnamefont{J.~A.} \bibnamefont{{Montgomery}}},
  \bibinfo{author}{\bibfnamefont{J.~N.} \bibnamefont{{Byrd}}},
  \bibinfo{author}{\bibfnamefont{H.~H.} \bibnamefont{{Michels}}},
  \bibnamefont{and}
  \bibinfo{author}{\bibfnamefont{R.}~\bibnamefont{{C{\^o}t{\'e}}}},
  \bibinfo{journal}{Chem. Phys. Lett.} \textbf{\bibinfo{volume}{542}},
  \bibinfo{pages}{138} (\bibinfo{year}{2012}), \eprint{1206.1304}.

\bibitem[{\citenamefont{{Kaur} et~al.}(2015)\citenamefont{{Kaur}, {Nandy},
  {Arora}, and {Sahoo}}}]{2015PhRvA..91a2705K}
\bibinfo{author}{\bibfnamefont{J.}~\bibnamefont{{Kaur}}},
  \bibinfo{author}{\bibfnamefont{D.~K.} \bibnamefont{{Nandy}}},
  \bibinfo{author}{\bibfnamefont{B.}~\bibnamefont{{Arora}}}, \bibnamefont{and}
  \bibinfo{author}{\bibfnamefont{B.~K.} \bibnamefont{{Sahoo}}},
  \bibinfo{journal}{\pra} \textbf{\bibinfo{volume}{91}}, \bibinfo{eid}{012705}
  (\bibinfo{year}{2015}).

\bibitem[{\citenamefont{{Chattopadhyay}
  et~al.}(2014)\citenamefont{{Chattopadhyay}, {Mani}, and
  {Angom}}}]{2014PhRvA..89b2506C}
\bibinfo{author}{\bibfnamefont{S.}~\bibnamefont{{Chattopadhyay}}},
  \bibinfo{author}{\bibfnamefont{B.~K.} \bibnamefont{{Mani}}},
  \bibnamefont{and} \bibinfo{author}{\bibfnamefont{D.}~\bibnamefont{{Angom}}},
  \bibinfo{journal}{\pra} \textbf{\bibinfo{volume}{89}}, \bibinfo{eid}{022506}
  (\bibinfo{year}{2014}).

\bibitem[{\citenamefont{Kramida and \uppercase{NIST ASD} Team}(2014)}]{NIST}
\bibinfo{author}{\bibfnamefont{R.~Y. R.~J.} \bibnamefont{Kramida},
  \bibfnamefont{A.}} \bibnamefont{and}
  \bibinfo{author}{\bibnamefont{\uppercase{NIST ASD} Team}},
  \emph{\bibinfo{title}{{NIST} atomic spectra database (v5.2) [online]}}
  (\bibinfo{year}{2014}), \bibinfo{note}{http://physics.nist.gov/asd}.

\bibitem[{\citenamefont{{Napolitano} et~al.}(1994)\citenamefont{{Napolitano},
  {Weiner}, {Williams}, and {Julienne}}}]{1994PhRvL..73.1352N}
\bibinfo{author}{\bibfnamefont{R.}~\bibnamefont{{Napolitano}}},
  \bibinfo{author}{\bibfnamefont{J.}~\bibnamefont{{Weiner}}},
  \bibinfo{author}{\bibfnamefont{C.~J.} \bibnamefont{{Williams}}},
  \bibnamefont{and} \bibinfo{author}{\bibfnamefont{P.~S.}
  \bibnamefont{{Julienne}}}, \bibinfo{journal}{\prl}
  \textbf{\bibinfo{volume}{73}}, \bibinfo{pages}{1352} (\bibinfo{year}{1994}).

\bibitem[{\citenamefont{{Juarros} et~al.}(2006)\citenamefont{{Juarros},
  {Kirby}, and {C{\^o}t{\'e}}}}]{2006JPhB...39S.965J}
\bibinfo{author}{\bibfnamefont{E.}~\bibnamefont{{Juarros}}},
  \bibinfo{author}{\bibfnamefont{K.}~\bibnamefont{{Kirby}}}, \bibnamefont{and}
  \bibinfo{author}{\bibfnamefont{R.}~\bibnamefont{{C{\^o}t{\'e}}}},
  \bibinfo{journal}{J. Phys. B} \textbf{\bibinfo{volume}{39}},
  \bibinfo{pages}{965} (\bibinfo{year}{2006}).

\bibitem[{\citenamefont{{Herzberg}}(1950)}]{Herzberg}
\bibinfo{author}{\bibfnamefont{G.}~\bibnamefont{{Herzberg}}},
  \emph{\bibinfo{title}{{Molecular spectra and molecular structure. Vol.1:
  Spectra of diatomic molecules}}} (\bibinfo{year}{1950}).

\bibitem[{\citenamefont{{Kokoouline} et~al.}(1999)\citenamefont{{Kokoouline},
  {Dulieu}, {Kosloff}, and {Masnou-Seeuws}}}]{1999JChPh.110.9865K}
\bibinfo{author}{\bibfnamefont{V.}~\bibnamefont{{Kokoouline}}},
  \bibinfo{author}{\bibfnamefont{O.}~\bibnamefont{{Dulieu}}},
  \bibinfo{author}{\bibfnamefont{R.}~\bibnamefont{{Kosloff}}},
  \bibnamefont{and}
  \bibinfo{author}{\bibfnamefont{F.}~\bibnamefont{{Masnou-Seeuws}}},
  \bibinfo{journal}{J. Chem. Phys.} \textbf{\bibinfo{volume}{110}},
  \bibinfo{pages}{9865} (\bibinfo{year}{1999}).

\bibitem[{\citenamefont{Johnson}(1978)}]{ren_numerov}
\bibinfo{author}{\bibfnamefont{B.~R.} \bibnamefont{Johnson}},
  \bibinfo{journal}{J. Chem. Phys.} \textbf{\bibinfo{volume}{69}},
  \bibinfo{pages}{4678} (\bibinfo{year}{1978}).

\bibitem[{\citenamefont{Demtr{\"o}der}(2007)}]{demtroder2007laserspektroskopie}
\bibinfo{author}{\bibfnamefont{W.}~\bibnamefont{Demtr{\"o}der}},
  \emph{\bibinfo{title}{Laserspektroskopie: Grundlagen und Techniken}}
  (\bibinfo{publisher}{Springer-Verlag}, \bibinfo{year}{2007}).

\bibitem[{\citenamefont{{Pellegrini} et~al.}(2008)\citenamefont{{Pellegrini},
  {Gacesa}, and {C{\^o}t{\'e}}}}]{2008PhRvL.101e3201P}
\bibinfo{author}{\bibfnamefont{P.}~\bibnamefont{{Pellegrini}}},
  \bibinfo{author}{\bibfnamefont{M.}~\bibnamefont{{Gacesa}}}, \bibnamefont{and}
  \bibinfo{author}{\bibfnamefont{R.}~\bibnamefont{{C{\^o}t{\'e}}}},
  \bibinfo{journal}{\prl} \textbf{\bibinfo{volume}{101}}, \bibinfo{eid}{053201}
  (\bibinfo{year}{2008}).

\end{thebibliography}

\end{document}